\documentclass[twocolumn]{aastex63}
\usepackage{amsmath,here,afterpage}
\usepackage{longtable}
\usepackage{multirow,comment}
\usepackage[pagewise]{lineno}

\received{}
\revised{}
\accepted{}
\submitjournal{ApJ}

\shorttitle{Atomic Carbon in the Central Molecular Zone of the Milky Way}
\shortauthors{Tanaka et al.}
\usepackage{float}
\newcommand\pcc{\ifmmode\mathrm{cm^{-3}}\else{$\mathrm{cm^{-3}}$}\fi}
\newcommand\psc{\mathrm{cm^{-2}}}
\newcommand\kelvin{\ifmmode\mathrm{ K}\else K\fi}

\newcommand\kmps{\ifmmode\mathrm{km\,s^{-1}}\else$\mathrm{km\,s^{-1}}$\fi}
\newcommand\erg{\mathrm{erg}}
\newcommand\ergs{\mathrm{ergs}}
\newcommand\pc{\mathrm{pc}}

\newcommand\Msol{\ifmmode{M_\odot}\else${M_\odot}$\fi}

\newcommand\kB{{k_{\rm B}}}

\newcommand\Tkin{\ifmmode{T_{\rm kin}}\else{$T_{\mathrm{kin}}$}\fi}
\newcommand\Tex{\ifmmode{T_{\rm ex}}\else{$T_{\mathrm{ex}}$}\fi}
\newcommand\Tturb{\ifmmode{T_{\rm turb}}\else{$T_{\mathrm{turb}}$}\fi}
\newcommand\Td{\ifmmode{T_{\mathrm{d}}\else{$T_{\mathrm{d}}$}\fi}}
\newcommand\Tv{\ifmmode{T_v}\else{${T_v}$}\fi}

\newcommand\Trot{\ifmmode{T_{\rm rot}}\else{$T_{\rm rot}$}\fi}

\newcommand\nH{\ifmmode{n_{\rm H}}\else${n_{\rm H}}$\fi}
\newcommand\nHH{\ifmmode{n_{\rm H_2}}\else{$n_{\rm H_2}$}\fi}
\newcommand\NHH{\ifmmode{N_{\rm H_2}}\else{$N_{\rm H_2}$}\fi}
\newcommand\NHHavg{\ifmmode{\left\langle N_{\rm H_2}\right\rangle}\else{$\left\langle N_{\rm H_2}\right\rangle$}\fi}
\newcommand\nHHavg{\ifmmode{\left\langle n_{\rm H_2}\right\rangle}\else{$\left\langle n_{\rm H_2}\right\rangle$}\fi}
\newcommand\ncrit{\ifmmode{n_{\rm crit}}\else{$n_{\rm crit}$}\fi}

\newcommand\vlsr{\ifmmode{v_{\rm LSR}}\else${v_{\rm LSR}}$\fi}
\newcommand\Mvt{\ifmmode{M_{\rm VT}}\else${M_{\rm VT}}$\fi}
\newcommand\Mhnc{\ifmmode{M_{\rm HNC}}\else${M_{\rm HNC}}$\fi}
\newcommand\Tmb{\ifmmode{T_{\rm MB}}\else{$T_{\mathrm{MB}}$}\fi}

\newcommand\dv {\ifmmode{\rm d}v\else${\rm d}v$\fi}
\newcommand\Eu{\ifmmode{E_{\rm u}}\else{$E_\mathrm{u}$}\fi}

\newcommand\vcol{\ifmmode{v_{\rm col}}\else{$v_{\rm col}$}\fi}
\newcommand\CO{\ifmmode\mathrm{CO}\else$\mathrm{CO}$\fi}
\newcommand\CN{\ifmmode\mathrm{CN}\else$\mathrm{CN}$\fi}

\newcommand\HOCp{\ifmmode{\rm HOC^+}\else${\mathrm HOC^+}$\fi}
\newcommand\CS{\ifmmode{\rm CS}\else${\mathrm CS}$\fi}
\newcommand\SiO{\ifmmode{\rm SiO}\else${\mathrm SiO}$\fi}

\newcommand\HHHp{\ifmmode{\rm H_3^+}\else{${\rm H_3^+}$}\fi}

\newcommand\HCN{\ifmmode{\rm HCN}\else{HCN}\fi}
\newcommand\HCOp{\ifmmode{\rm HCO^+}\else{$\mathrm{HCO^+}$}\fi}
\newcommand\HCNt{\ifmmode{\rm H{^{13}C}N}\else{$\mathrm{H{^{13}C}N}$}\fi}
\newcommand\HNC{\ifmmode{\rm HNC}\else{$\mathrm{HNC}$}\fi}
\newcommand\HCCCN{\ifmmode{\rm HC_3N}\else{$\mathrm{HC_3N}$}\fi}
\newcommand\HCOpt{\ifmmode{\rm H^{13}CO^+}\else{$\mathrm{H^{13}CO^+}$}\fi}

\newcommand\Cn{\ifmmode {\rm C^0}\else $\mathrm{C^0}$\fi}
\newcommand\Cp{{\rm C^+}}
\newcommand\COt{\ifmmode{\rm {^{13}CO}}\else{$\mathrm{^{13}CO}$}\fi}
\newcommand\Ct{\ifmmode{\rm {^{13}C}}\else{$\mathrm{^{13}C}$}\fi}
\newcommand\Ctw{\ifmmode{\rm {^{12}C}}\else{$\mathrm{^{12}C}$}\fi}
\newcommand\ammonia{\ifmmode{\rm NH_3}\else{$\rm NH_3$}}

\newcommand\NNHp{\ifmmode{\rm N_2H^+}\else{$\mathrm{N_2H^+}$}\fi}
\newcommand\HHCS{\ifmmode{\rm H_2CS}\else{$\mathrm{H_2CS}$}\fi}
\newcommand\CtS{\ifmmode{\rm ^{13}CS}\else{$\mathrm{^{13}CS}$}\fi}
\newcommand\OCS{\ifmmode{\rm OCS}\else{$\mathrm{OCS}$}\fi}
\newcommand\methanol{\ifmmode{\rm CH_3OH}\else{$\mathrm{CH_3OH}$}\fi}
\newcommand\pformaldehyde{\ifmmode{p\mbox{-}\rm H_2CO}\else{$p$-$\mathrm{H_2CO}$}\fi}
\newcommand\formaldehyde{\ifmmode{\rm H_2CO}\else{$\mathrm{H_2CO}$}\fi}
\newcommand\pfx{\ifmmode{3_{21}--2_{20}}\else{$3_{21}--2_{20}$}\fi}
\newcommand\pfy{\ifmmode{3_{22}--2_{21}}\else{$3_{22}--2_{21}$}\fi}
\newcommand\HHO{\ifmmode{\rm H_2O}\else{$\mathrm{H_2O}$}\fi}

\newcommand\JJ[2]{\ifmmode{\mbox{{\it J}={#1}\mbox{--}{#2}}}\else{{\it J}={#1}--{#2}}\fi}
\newcommand\JJx[2]{\ifmmode{\mbox{{#1}\mbox{--}{#2}}}\else{{#1}--{#2}}\fi}
\newcommand\NN{{\it N}}
\newcommand\JK[4]{\ifmmode{{J_K}=#1_{#2}\mbox{--}#3_{#4}}\else{${\it J_K}=#1_{#2}\mbox{--}#3_{#4}$}\fi}
\newcommand\JN[4]{\mbox{{\it $J_N$}=$#1_{#2}$\mbox{--}$#3_{#4}$}}
\newcommand\NJ[4]{\mbox{{\it $N_J$}=$#1_{#2}$\mbox{--}$#3_{#4}$}}
\newcommand\CIa{\ifmmode{^3}P_1\mbox{--}{^3}P_0\else${^3}P_1\mbox{--}{^3}P_0$\fi}
\newcommand\CIb{\ifmmode{^3}P_2\mbox{--}{^3}P_1\else${^3}P_2\mbox{--}{^3}P_1$\fi}

\newcommand\gl{\ifmmode l\else{\it l}\fi}
\newcommand\gb{\ifmmode b\else{\it b}\fi}

\newcommand\sgras{$\mathrm{Sgr A^{*}}$}

\newcommand\CLb{CO$-0.30$$-0.07$}
\newcommand\theObj\CLb

\newcommand\Dv{\ifmmode{\Delta v}\else{$\Delta v$}\fi}
\newcommand\Dvheat{\ifmmode{\Delta v_{\rm heat}}\else{$\Delta v_{\rm heat}$}\fi}
\newcommand\Sv{\ifmmode{\sigma v}\else{$\sigma v$}\fi}
\newcommand\vc{\ifmmode{\left<v\right>}\else{$\left<v\right>$}\fi}

\newcommand{\avir}{\ifmmode{\alpha_{\rm vir}}\else{$\alpha_{\rm vir}$}\fi}
\newcommand{\ntild}{\ifmmode{n^*}\else{$n^*$}\fi}

\newcommand\Sdust{\ifmmode{S_{500}}\else{$S_{500}$}\fi}
\newcommand\II[1]{\ifmmode{I_{#1}}\else{$I_{#1}$}\fi}
\newcommand\myvector[1]{\ifmmode{\mbox{\boldmath ${#1}$}}\else{\boldmath {${#1}$}}\fi}
\newcommand{\Rt}{\ifmmode{{R_{13}}}\else${R_{13}}$\fi}
\newcommand{\Iobs}{\ifmmode{I}\else${I}$\fi}
\newcommand{\Icalc}{\ifmmode{{F}\left(\pv\right)}\else${{F}\left(\pv\right)}$\fi}
\newcommand{\Icalci}{\ifmmode{{F}\left(\pv_{i}\right)}\else${{F}\left(\pv_{i}\right)}$\fi}
\newcommand{\Icalcv}{\ifmmode{\myvector{F}\left(\pv\right)}\else${\myvector{F}\left(\pv\right)}$\fi}
\newcommand{\xmol}{\ifmmode{{x_{\rm mol}}}\else${x_{\rm mol}}$\fi}
\newcommand{\xmolp}[1]{\ifmmode{{x_{\rm mol}\left(#1\right)}}\else${x_{\rm mol}\left(\mbox{#1}\right)}$\fi}
\newcommand{\ff}{\ifmmode{\Phi}\else${\Phi}$\fi}
\newcommand{\fff}{\ifmmode{\phi}\else${\phi}$\fi}
\newcommand{\fcal}{\ifmmode{f_{\rm cal}}\else${f_{\rm cal}}$\fi}

\newcommand\unit[1]{\ifmmode{\mbox{\boldmath $e$}_{#1}}\else{${\boldmath e}_{#1}$}\fi}

\newcommand{\Ea}{\ifmmode\epsilon^{\rm a}\else$\epsilon_{\rm a}$\fi}
\newcommand{\Em}{\ifmmode\epsilon\else$\epsilon$\fi}

\newcommand{\sa}{\ifmmode\sigma\else$\sigma$\fi}
\newcommand{\sm}{\ifmmode\sigma\else$\sigma$\fi}
\newcommand{\scal}{\ifmmode\sigma_{\rm cal}\else$\sigma_{\rm cal}$\fi}

\newcommand{\av}{\myvector{a}}

\newcommand{\pv}{\myvector{p}}
\newcommand{\pvi}{\myvector{p_i}}

\newcommand{\vv}{\myvector{v}}

\newcommand{\qv}{\myvector{q}}
\newcommand{\qvi}{\myvector{q_i}}

\newcommand{\PDF}[1]{{\ifmmode P(#1) \else $P(#1)$ \fi}}%{\ifmmode{P\left({#1}\right)}\else{$P\left({#1}\right)$}\fi}

\newcommand\RRx{\ifmmode R_{43}\else$R_{43}$\fi}
\newcommand\IIx{\ifmmode I_{13}\else$I_{13}$\fi}
\newcommand\Ihnc{\ifmmode I_{\HNC}\else$I_{\HCN}$\fi}
\newcommand\Ihcn{\ifmmode I_{\HCN 43}\else$I_{\HCN 43}$\fi}
\newcommand\Ihcnt{\ifmmode I_{\HCNt}\else$I_{\HCNt}$\fi}
\newcommand\Ihcccn{\ifmmode I_{\HCCCN}\else$I_{\HCCCN}$\fi}

\newcommand\Lhnc{\ifmmode L_{\HNC}\else$L_{\HCN}$\fi}
\newcommand\Lhcn{\ifmmode L_{\HCN 43}\else$L_{\HCN 43}$\fi}
\newcommand\Lhcnt{\ifmmode L_{\HCNt}\else$L_{\HCNt}$\fi}
\newcommand\Lhcccn{\ifmmode L_{\HCCCN}\else$L_{\HCCCN}$\fi}

\newcommand\dNdv{\ifmmode {\mathrm{d}N}/{\mathrm{d}v}\else${\mathrm{d}N}/{\mathrm{d}v}$\fi}
\newcommand\dNHdv{\ifmmode \frac{\mathrm{d}N_{\rm H_2}}{\mathrm{d}v}\else$\frac{\mathrm{d}N_{\rm H_2}}{\mathrm{d}v}$\fi}
\newcommand\Xdvdr{\ifmmode {{X}/{\frac{\mathrm{d}v}{\mathrm{d}r}}}\else{${X}/\frac{\mathrm{d}v}{\mathrm{d}r}$}\fi}

\newcommand\Nobs{\ifmmode{N_{\rm obs}}\else{$N_{\rm obs}$}\fi}
\newcommand\Nmol{\ifmmode{N_{\rm mol}}\else{$N_{\rm mol}$}\fi}
\newcommand\Np{\ifmmode{N_{p}}\else{$N_{p}$}\fi}
\newcommand\Nl{\ifmmode{N_{l}}\else{$N_{l}$}\fi}

\newcommand\dvdr{\ifmmode{{\mathrm d}v/{\mathrm d}r}\else{${\mathrm d}v/{\mathrm d}r$}\fi}

\newcommand\Mmag{\ifmmode M_\Phi\else$M_\Phi$\fi}
\newcommand\nth{\ifmmode {n_{\mathrm{th}}}\else$n_{\mathrm{th}}$\fi}
\newcommand\SFRff{\ifmmode {\mathrm{SFR_{ff}}}\else$\mathrm{SFR_{ff}}$\fi}

\newcommand\zCR{\ifmmode \zeta_{\mathrm{CR}}\else$\zeta_{\mathrm{CR}}$\fi}
\newcommand\xe{\ifmmode x_{\mathrm{e}}\else$x_{\mathrm{e}}$\fi}
\newcommand\Qc{\ifmmode Q_\mathrm{c} \else $Q_\mathrm{c}$\fi}
\newcommand\Qnc{\ifmmode Q_\mathrm{nc} \else $Q_\mathrm{nc}$\fi}

\newcommand\logt{\ifmmode \log_{10}\else $\log_{10}$\fi}
\newcommand\parsec{\ifmmode pc\else $\mathrm{pc}$\fi}
\newcommand\Prob[1]{\ifmmode \mathrm{Pr}\left(#1\right) \else $\mathrm{Pr}\left(#1\right)$ \fi}
\newcommand\avpv{\ifmmode \av\cdot\pv \else $\av\cdot\pv$\fi}

\newcommand\fsf{\ifmmode {f_{\mathrm{SF}}}\else${f_{\mathrm{SF}}}$\fi}
\newcommand\tff{\ifmmode {t_{\mathrm{ff}}}\else${t_{\mathrm{ff}}}$\fi}
\newcommand\epsff{\ifmmode {\epsilon^*_{\mathrm{ff}}}\else${\epsilon^*_{\mathrm{ff}}}$\fi}
\newcommand\etasf{\ifmmode {\eta_{\mathrm{SF}}}\else${\eta_{\mathrm{SF}}}$\fi}
\newcommand\etadense{\ifmmode {\eta_{\mathrm{dense}}}\else${\eta_{\mathrm{dense}}}$\fi}
\newcommand\ith{\ifmmode {{i}^{\mathrm{th}}}\else ${{i}^{\mathrm{th}}}$\fi}
\newcommand\Npar{\ifmmode {N_{\rm param}}\else ${N_{\mathrm{param}}}$\fi}

\newcommand\SigmaH{\ifmmode {\Sigma_{\rm H_2}}\else ${\Sigma_{\mathrm{H_2}}}$\fi}

\newcommand\Myr{\ifmmode {\rm Myr}\else Myr\fi}
\newcommand\rmaj{\ifmmode{r_{\mathrm{maj}}}\else${r_{\mathrm{maj}}}$\fi}
\newcommand\rmin{\ifmmode{r_{\mathrm{min}}}\else${r_{\mathrm{min}}}$\fi}

\newcommand\ttr{\ifmmode{\tau_{\mathrm{cl}}}\else $\tau_{\mathrm{cl}}$\fi}

\newcommand\CI{[\ion{C}{1}]}
\newcommand\CII{[\ion{C}{2}]}

\newcommand\Tci{\ifmmode T({\rm CI})\else $T({\rm CI})$\fi}
\newcommand\Tct{\ifmmode T(\COt)\else $T({\COt})$\fi}

\newcommand\dCI{\ifmmode {\mathrm{d}T}\else {$\mathrm{d}T$} \fi}
\newcommand\dCIx{\ifmmode {\mathrm{d}T_{1\text{--}0}}\else {$\mathrm{d}T_{1\text{--}0}$} \fi}
\newcommand\dCIy{\ifmmode {\mathrm{d}T_{2\text{--}1}}\else {$\mathrm{d}T_{2\text{--}1}$} \fi}
\newcommand\RCI{\ifmmode R\else $R$\fi}

\newcommand\Rco{\ifmmode R_{\rm 2\mathchar`-1/1\mathchar`-0}\else $R_{\rm 2\mathchar`-1/1\mathchar`-0}$\fi}

\newcommand\xCn{\NN(\Cn)/\NN(\CO)}

\newcommand\molH{\ifmmode\mathrm{H_2}\else$\mathrm{H_2}$\fi}
\newcommand\ps{\ifmmode\mathrm{s^{-1}}\else$\mathrm{yr^{-1}}$\fi}
\newcommand\yr{\ifmmode\mathrm{yr}\else$\mathrm{yr}$\fi}

\newcommand\Xcn{\xCn}
\newcommand\tr[1]{\ifmmode{{^t}#1}\else${{^t}#1}$\fi}

\newcommand\vp{\myvector{p}}
\newcommand\vq{\myvector{q}}

\newcommand\vqi{\myvector{q_i}}

\begin{document}
%%%%%%%%%%%%%%%%%%%%%%%%%%%%%%%%%%%%%%
%%\baselineskip=7mm %%!!!!!!!!!!!!!!!!!
%%%%%%%%%%%%%%%%%%%%%%%%%%%%%%%%%%%%%%

\title{Atomic Carbon in the Central Molecular Zone of the Milky Way : Possible Cosmic-ray Induced Chemistry or Time-Dependent Chemistry Associated with SNR Sagittarius A East}

\author{Kunihiko Tanaka}
\email{ktanaka@phys.keio.ac.jp}
\affil{Department of Physics, Faculty of Science and Technology, Keio University, 3-14-1 Hiyoshi, Yokohama, Kanagawa 223--8522 Japan}

\author{Makoto Nagai}
\affil{Advanced Technology Center, National Astronomical Observatory of Japan, 2-21-1 Osawa, Mitaka, Tokyo 181-8588, Japan}
\author{Kazuhisa Kamegai}
\affil{Astronomy Data Center, National Astronomical Observatory of Japan, 2-21-1 Osawa, Mitaka, Tokyo 181-8588, Japan}

%\keywords{Galaxy: center \object{Galactic Center}}

%somechange

%\linenumbers

\begin{abstract}

  Being one of the most abundant atomic/molecular species observed in dense molecular gas, atomic carbon ($\mathrm{C}^0$) is a potential good tracer of molecular gas mass in many chemical/physical environments, though the $\mathrm{C^0}$ abundance variation outside the Galactic disk region is yet to be fully known.
This paper presents a wide-field 500 GHz [\ion{C}{1}] map of the Galactic central molecular zone (CMZ) obtained with the ASTE 10-m telescope.
Principal component analysis and non-LTE multi-transition analysis
have shown that the [\ion{C}{1}] emission predominantly originates from the low-excitation gas component with a 20--50 K temperature and $\sim 10^3\ \mathrm{cm}^{-3}$ density, whereas $\mathrm{C^0}$ abundance is likely suppressed in the high-excitation gas component.
The average $N(\mathrm{C}^0)$/$N(\mathrm{CO})$ abundance ratio in the CMZ is 0.3--0.4, which is 2--3 times that in the Galactic disk.  The $N(\mathrm{C}^0)$/$N(\mathrm{CO})$ ratio increases to 0.7 in the innermost 10 pc region and to $\sim2$ in the circumnuclear disk. 
We discovered $\mathrm{C^0}$-rich regions distributed in a ring-shape encircling the supernova remnant (SNR) Sgr~A~east, indicative that the $\mathrm{C}^0$-enrichment in the central 10 pc region is a consequence of a molecular cloud--SNR interaction.
In the 15 atom/molecules included in principal component analysis (PCA),  CN is the only other species that increases in the [\ion{C}{1}]-bright ring.
The origin of the [\ion{C}{1}]-bright ring is likely a cosmic-ray dominated region created by low-energy cosmic-ray particles accelerated by Sgr~A~east  
or primitive molecular gas collected by the SNR in which the $\mathrm{C}^0$-to-CO conversion has not reached the equilibrium.

\end{abstract}

%%%%%%%%%%%%%%%%%%%%%%%%%%%%%%%%%%
%
% INTRODUCTION
%
%%%%%%%%%%%%%%%%%%%%%%%%%%%%%%%%%%

\section{INTRODUCTION\label{section:introduction}}

Atomic carbon (\Cn) is one of the most abundant atomic/molecular species observable in interstellar molecular clouds,
whose measured relative abundance to CO ranges from 0.1 to 10 in Galactic and extragalactic sources \citep{White1994,Fixsen1999,Maezawa1999,Israel2002,Hitschfeld2008,Tanaka2011,Izumi2020}.
{Despite the early prediction that the submillimeter \CI\ forbidden lines\footnote{{Throughout this paper, we denote the forbidden transitions from atomic carbon by the symbol `\CI', while `\Cn' indicates the carbon atom itself.}} mainly arise from a thin layer in the photodissociation region (PDR) developed at the molecular cloud surface} \citep{Tielens1985a,Tielens1985b,Hollenbach1991}, observations have proved that the \CI\ emission is coextensive with low-$J$ \COt\ lines over broad spatial scales from 0.1 pc to the entire-cloud scale \citep{Plume1994,Ikeda1999,Shimajiri2013}.
This rich abundance and ubiquitousness make \Cn\ a potential good gas mass tracer alternative to or even better than CO lines in the distant universe or high cosmic-ray (CR) flux environments \citep{Bisbas2017,Papadopoulos2018,Bourne2019}.

The \Cn\ abundance in molecular clouds varies over more than an order of magnitude depending on the environment.
The standard \Xcn\ abundance ratio in the Galactic disk region is 0.1--0.2 (\citealt{Maezawa1999,Ikeda2002,Kamegai2003,Sakai2006}; see also \S\ref{discussion:overall}), whereas highly enhanced \Xcn\ ratios of a few to 10 are observed for strong starburst galaxies, Seyfert, and (U)LIRGs \citep{Israel2002,Krips2016,Miyamoto2018,Izumi2020}.
Spots with elevated \Cn-abundances are also detected in the star-forming ring of NGC613 \citep{Miyamoto2018} and the root point of the bipolar CO outflow of the NGC253 \citep{Krips2016}, as well as in Galactic dark clouds \citep{Maezawa1999} and giant molecular clouds (GMCs) interacting with supernova remnants (SNRs)  \citep{White1994,Arikawa1999}.
In contrast, a \CI-faint region without detectable \CI~\CIa\ emission was recently discovered in the merging galaxy NGC6052 \
\citep{Michiyama2020}.

The central 200 pc region of the Galaxy, or the central molecular zone (CMZ), is the largest \Cn-rich region in the Galaxy, whose overall \xCn\ abundance ratio is approximately 2--3 times the Galactic disk value \citep{Jaffe1996,Ohja2001,Martin2004,Tanaka2011,Garcia2016}.
The \CI/\COt\ intensity ratio further increases by a factor of $>2$ within a 6 pc radius from the nucleus object, \sgras \citep{Ohja2001,Tanaka2011}.
In \cite{Tanaka2011},
we reported the discovery of the \CI-enhanced region in the eastern half of the Sgr~A complex, the closest GMC complex to the Galactic nucleus. 
The Sgr~A \CI-enhanced region comprises the 50-\kmps\ cloud, the {circumnuclear} disk (CND), and the clump CO{$0.02$}{$-0.02$} (hereafter CO$0.02$), whose \CI/\COt\ intensity ratios are $>0.8$; in particular, the CND and CO$0.02$ have the \CI/\COt\ intensity ratios $\sim 5$ times the CMZ average,    
though it remained undetermined whether their highly enhanced intensity ratios translate into high \Cn\ abundances or high \COt\ excitation temperatures.
Increasing gradients of the \CI/CO intensity ratio toward smaller galactic radii are commonly observed in galactic central regions \citep{Miyamoto2018,Salak2019,Izumi2020,Saito2020}, indicative that the \Cn-rich state is associated with the harsh environment of the galactic central regions characterized by intense cosmic-ray (CR) and X-ray fields, fast turbulent velocities, and short cloud lifetimes, as predicted by theories  \citep{Suzuki1992,Papadopoulos2004,Harada2019,Meijerink2005,Meijerink2007,Flower1994,Boger2005,Papadopoulos2010,Bisbas2015,Bisbas2017,Papadopoulos2018,Mitchell1984,Hollenbach1989}.

This paper presents a wide-field \CI~\CIa\ map of the CMZ with a $\sim 1\,\pc$ resolution, along with  CN~$N_J$=$1_{3/2}$--$0_{1/2}$ and CO~\JJ{3}{2}\ maps of the Sgr~A complex.
%The new \CI\ observation almost entirely covers the major three GMC complexes in the CMZ;
The new \CI\ observation covers almost the entire CMZ, including the Sgr~C complex and the southeastern extension of the Sgr~B2 complex, which were out of coverage of the near-Galactic plane mapping by \cite{Tanaka2011}. %, are within the field of view of the new observations.
Using the newly obtained data and multiline data compiled from literature, we attempt to characterize the physical and chemical properties of the \CI-emitting region by means of principal component analysis (PCA) and accurate non-LTE excitation analysis.
%We also investigate the spatial variation of the \Cn\ abundance based on accurate measurements of the physical conditions of the \CI-emitting region. 
The rest of this paper is structured as follows.  The next section
(\S\ref{section:observation}) describes the observation
performed with the ASTE 10-m and Nobeyama Radio Observatory (NRO) 45-m telescopes.  
The \CI~\CIa, \COt~\JJ{1}{0}, and CN $N_J$=$1_{3/2}$--$0_{1/2}$ data are presented in
Section \ref{section:results}, using which we perform a comparison with dense gas distribution, measurements of the {\NN(\Cn)/\NN(\CO)} abundance ratio, and PCA.
We explore the origin of the \Cn-rich state in the CMZ in
Section \ref{section:discussion}.  Section \ref{section:summary}
summarizes the main results.
We use the distance to the Galactic center of 8.18 kpc \citep{GravityCollaboration2019} in all analyses in this paper.

\section{OBSERVATION and DATA REDUCTION\label{section:observation}}\subsection{ASTE [CI]~\CIa\ and \COt~\JJ{3}{2}\ Observations\label{observation:ASTE}}

We conducted wide-field mapping observations of the CMZ in
the \CI\ \CIa\ (492.1607 GHz) line and mapping of the Sgr A complex in the 
\COt\ \JJ{3}{2}\ (345.7960 GHz) line,
using the Atacama Submillimeter Telescope Experiment 10-m telescope
\citep[ASTE;][]{Ezawa2004}.  The observations were conducted in three
semesters, from October to November 2010, from May to July 2015, and
October 2016.  As we published a separate article for the 2010
observations \citep{Tanaka2011}, we describe the second and third
observations in the following. % part of this section.

The new \CI\ observation fully covers the Sgr A, B2, and C complexes in addition to the near--Galactic plane region observed in the 2010 observations.  The
observations were conducted in 9 and 11 nights in the 2015 and 2016
semesters, respectively, yielding a total on-source integration time of
approximately 30 hours after screening low-quality data. 
The 500 GHz \CI\ data were obtained using the ALMA band-8 QM receiver in the dual-polarization mode.  
The typical system noise temperatures during the observation run were 2000--3000 K per
polarization.  
The WHSF correlator system \citep{Iguchi2008} and the MAC digital spectrometer were used
as the backend.  The WHSF was operated in the 2048-MHz bandwidth mode,
which provided a 1200\ \kmps\ velocity coverage and 0.6\ \kmps\
channel separation at 500 GHz.  The HANNING function was chosen as the
window function.  The MAC spectrometer was used as a backup when the
WHSF had a stability problem.  As the spectral coverage of a single
array of the MAC in the wide-band mode ($\sim 300\ \kmps$ at 500 GHz)
is narrower than the overall velocity range of the CMZ ($\sim
500\ \kmps$), we composed full-bandwidth spectra by combining two arrays that are
configured to cover a 500\ \kmps\ range with a 100\ \kmps\ overlap.

The \COt\ \JJ{3}{2}\ observation was conducted as a backup observation when the atmospheric conditions were insufficient for 500 GHz observations.
The mapping was limited to the $15'\times10'$ area of the Sgr
A complex containing the 50-\kmps\ cloud, 20-\kmps\ cloud, and the
CND.  The DASH345 receiver and the MAC digital spectrometer were used
as the receiver frontend and backend, respectively.  The typical system
noise temperature was 500--1000 K.

The mapping observations were conducted with multiple on-the-fly (OTF)
scans each covering $10'\times10'$ or smaller area both in the X- and
Y-directions, i.e., in the directions of the Galactic longitude and
latitude, respectively, except for 
that the Galactic western half of the \COt\ data lacks X-scan maps due to shortage of observation time. 
The off-position spectra were taken at the position
$(\gl, \gb)$=$(1^\circ, -1^\circ)$, where no \CI\ \CIa\ emission was
detected above the noise level in the 2010 observation.  The antenna
pointing accuracy was maintained by making 5-point observations of the
CO~\JJ{4}{3}\ or \JJ{3}{2}\ emission toward V1427~Aql more than once every 1 hour.  
The residual antenna offset after the pointing measurements was less than
$5''$.  The antenna temperature was calibrated with the standard
chopper-wheel method, at least once every 20 minutes during the
observations.

The spectral data were reduced into a position--position--velocity (PPV)
data cube by utilizing the NOSTAR package developed by the NRO.  Up to
third-degree polynomial fitting was applied for spectral baseline
subtraction. 
A few spectra had baseline noise of a more complex shape than a third-degree polynomial;  for those spectra, the fitting velocity ranges were chosen so that the $-150$ to 150\ \kmps\ spectral range becomes flat.
A Gaussian-tapered Bessel function was used as the
convolution kernel to resample the OTF-sampled spectra into a PPV
data cube with an $8.5''\times8.5''\times2~\kmps$ PPV grid spacing.  
The effective {spatial resolutions} are $20''$ and $26''$ for the \CI\ and the \COt\ maps, respectively
 including the beam widening of $\sim 30\%$ by the convolution kernel.
The PLAIT algorithm \citep{Emerson1988} was applied to remove scanning noise for the 
regions where both X- and Y-scan maps were obtained.
The Galactic western half of the \COt\ \JJ{3}{2}\ map was obtained only with the Y-scans,
for which we applied the PRESS \citep{Sofue1979} algorithm to remove the scanning effect.

The intensity scale was first corrected for main-beam efficiency
and image rejection ratio by applying the scaling factor measured
with the calibrator measurements in previous observations \citep{Tanaka2011,Tanaka2018b}.
For the \CI\ data, we further checked the consistency with the 2010
data by comparing the intensities at the peak position of the
50-\kmps\ cloud, and found an increase of 17\% in the intensity from
the 2010 data to the newly taken data.  Finally, we coadded the 2010
data and the newly taken data into one data cube, after correcting the
intensity scale of the 2015 and 2016 data for the above intensity mismatch.

The \CI\ \CIa\ map is presented in the next section (\S\ref{section:results:oveall}).
The full \COt\ \JJ{3}{2}\ map is presented in Appendix \ref{appendix:OtherMaps}.
The \CI\ spectra in the Sgr~B2 and C complexes are relatively heavily affected by the residual of the spectral baseline subtraction.
The 180-pc ring, whose emission mainly appears in the $\pm (150$--$200)$\ \kmps\ velocity ranges, overlaps with this baseline noise, causing the intensity measurement to be unreliable for the 180-pc ring.
Therefore, we exclude the 180-pc ring from the analysis in this paper.

\subsection{NRO Observations\label{observation:NRO}}

The CN~\NJ{1}{3/2}{0}{1/2}\ (113.5 GHz) map of the Sgr A complex was taken using the Nobeyama Radio
Observatory (NRO) 45-m telescope, along with the \methanol~$0_0$--$1_{-1}$E (108.9 GHz) map simultaneously obtained in the other sideband. 
The observations were conducted in January and February 2010 using the 25-beam array receiver system
BEARS \citep{Sunada2000}.
The NRO 45-m observations covered the $30'\times 20'$ area containing
the entire Sgr A complex, M$0.11$$-0.08$, and G0.253+0.016.

We operated the digital backend in the wide-band mode with a channel
separation of 0.5 MHz and a total bandwidth of 512 MHz, which
corresponds to a velocity channel separation of 1.4\ \kmps\ and a
velocity coverage of 1400\ \kmps\ at 110 GHz. 
The target region was mapped by performing 6 square regions each covering
a $10'\times 10'$ area, both in the $X$- and $Y$-directions.  Antenna
pointing accuracy was maintained within $3''$ by observing the SiO
$\JJ{1}{0},\ v=1,2$ maser lines toward VX-Sgr.

The data were converted into $\gl$-$\gb$-\vlsr\ data cubes with a
$20.55''\times20.55''\times2\ \kmps$ grid using the same procedure
as that in the ASTE data reduction, except for additional
correction for variation in the sideband ratios among the 25 receiver
beams.
We used the correction factors measured by the observatory at 115 GHz in the upper sideband mode.
We applied the same scaling factors for the \methanol\ line simultaneously observed in the lower sideband, as the scaling factors for the lower sideband were not provided by the observatory;
hence, the intensity calibration of the \methanol\ line is likely to include large uncertainty of a few 10\%.
For $\eta_{\rm MB}$ of the CN and \methanol\ lines, we used the values at 115~GHz and 109~GHz measured by the observatory, i.e., 0.39 and 0.45, respectively.
The effective beam size is {$20''$} including the beam widening of $\sim 30\%$ 
by the kernel convolution applied in the OTF data reduction.
The full CN and \methanol\ maps are presented in Appendix \ref{appendix:OtherMaps}\ along with the ASTE \COt\ \JJ{3}{2}\ data.

\subsection{Hyperfine Deconvolution of the CN Line}
{
The CN~\NJ{1}{3/2}{0}{1/2}\ line consists of five hyperfine components distributed in the frequency interval from $-2.8$ to +29.3 MHz  around the main component.
This frequency interval corresponds to the velocity interval of $-77.8$ to +7.5~\kmps\ at 113 GHz, which is wider than the typical velocity widths in the CMZ ($\sim10\ \kmps$) and comparable to the cloud--to--cloud dispersion of the centroid velocities in the Sgr A complex ($\sim 100\ \kmps$).}
The intensity ratio of the strongest ($F=5/2$--$3/2$; 113.49097 GHz) to the next strongest component ($F=3/2$--$1/2$; 113.48812 GHz) is 2.65, indicating that the contamination from the satellite lines is non-negligible when we compare the CN distribution with \CI\ and other lines in the PPV space.

We removed the hyperfine satellite lines by applying a spectral deconvolution using the Fourier quotient method.
The details of the hyperfine deconvolution are described in Appendix \ref{appendix:CNdeconv}.
In all analyses in this paper, we use the hyperfine-deconvolved CN cube, unless otherwise stated.

\section{RESULTS AND ANALYSIS\label{section:results}}

\begin{figure*}[p]
  \epsscale{1.}
  \begin{center}
    \begin{tabular}{c!{\hspace{2cm}}c}
    \includegraphics[angle=90,height=21cm]{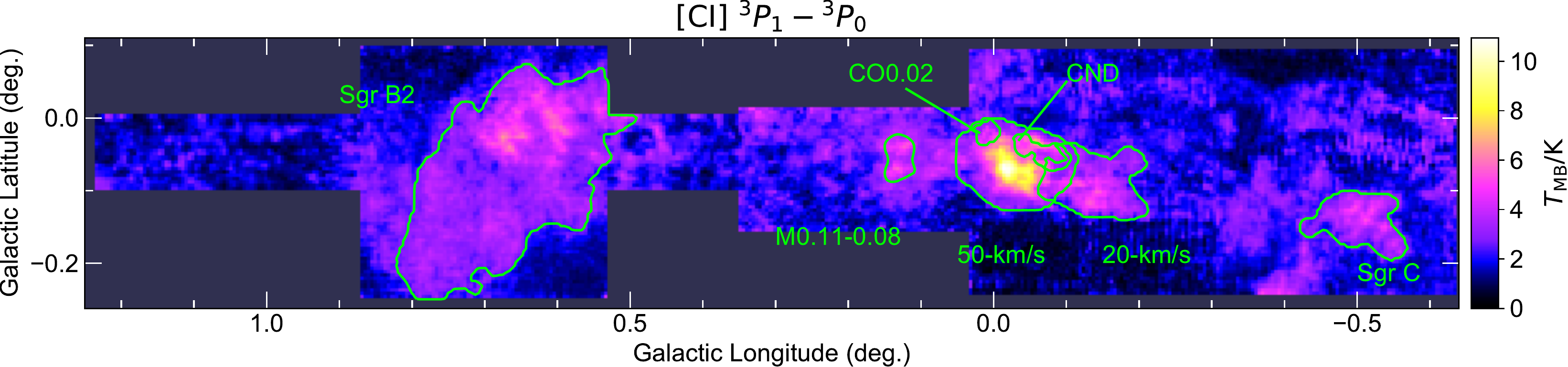}  &
    \includegraphics[angle=90,height=21cm]{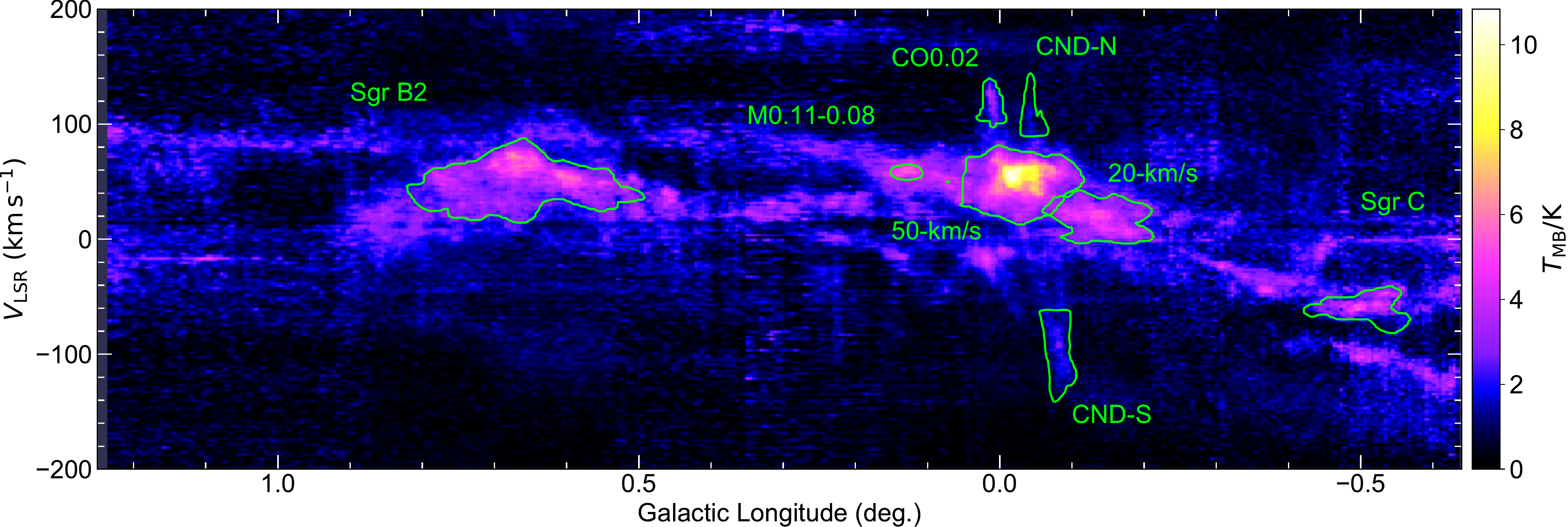}
    \end{tabular}
\end{center}
  \caption{Peak-intensity maps of the \CI\ \CIa\ data obtained with the ASTE 500-GHz band observations on the \gl--\gb\ (top) and \gl--\vlsr\ (bottom) planes.
    The peak intensities are calculated using voxels that are 5-channel binned along the axes perpendicular to the respective projected plane.
    \label{fig:ppmap}}
\end{figure*}

\begin{figure}[ttt]
%\ifdraft
%\epsscale{0.6}
%\else
\epsscale{1.}
%\fi
\plotone{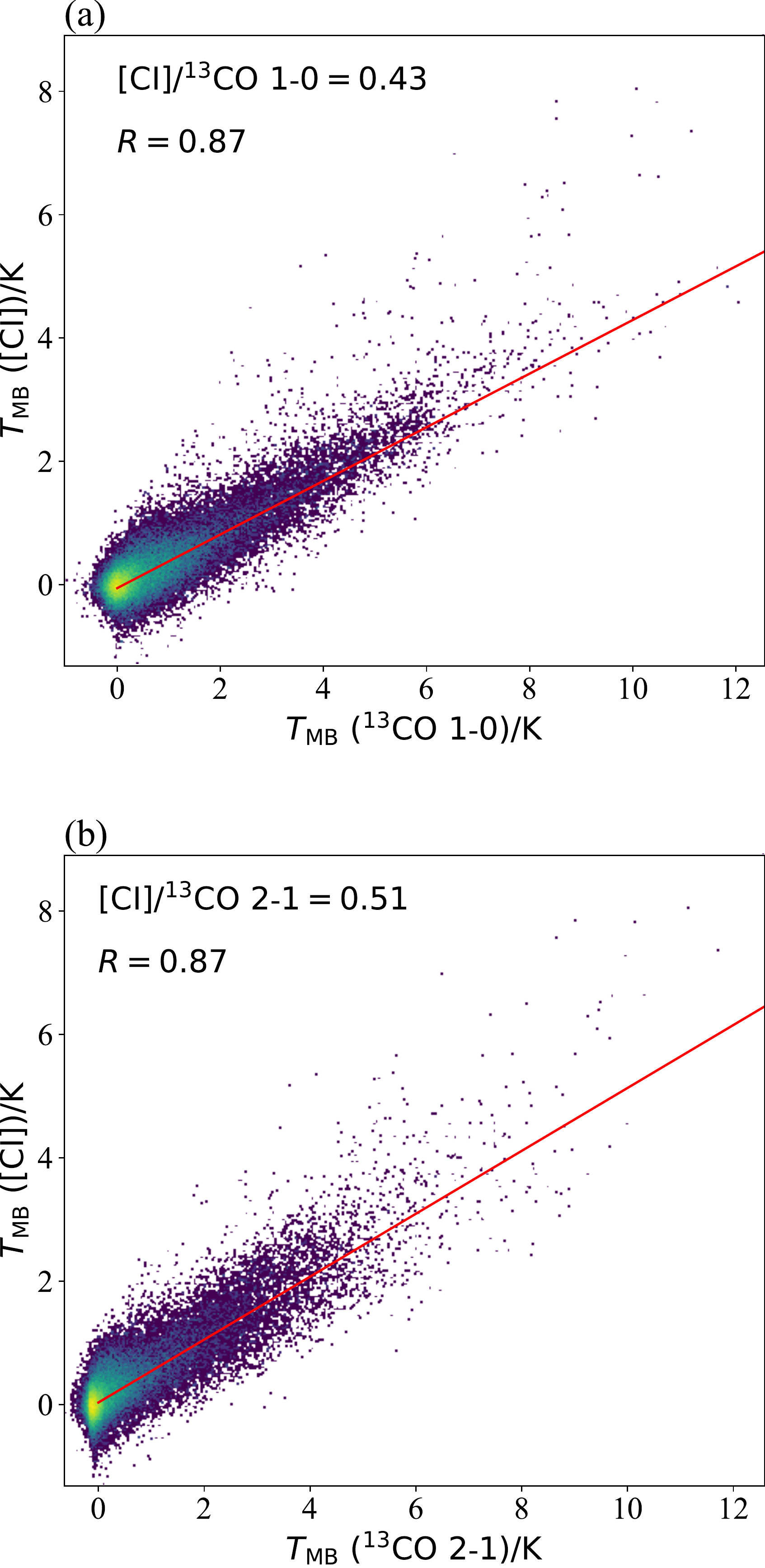}
\caption{Scatter plots of \CI\ \CIa\ vs. (a) \COt\ \JJ{1}{0}\ and (b) \COt \JJ{2}{1}. 
  The red lines are the results of robust fitting using the bi-weight function.  
  The overall intensity ratios and correlation coefficients ($R$) are shown in each panel.
  \label{fig:13CO_CI_scatterPlot}}
\end{figure}

\subsection{\CI-enhanced Regions\label{section:results:oveall}\label{section:results:CIbrightRegions}}
Figure \ref{fig:ppmap} shows the peak intensity map of the
\CI\ \CIa\ emission projected on \gl--\gb\ and \gl--\vlsr\ planes. 
The approximate boundaries of major GMCs and GMC complexes (50-\kmps\ cloud, 20-\kmps\ cloud, Sgr~B2, Sgr~C, CND, M0.11$-0.08$) and the cloud CO0.02$-0.02$ (hereafter CO0.02) are overlaid on the maps.

Figure \ref{fig:13CO_CI_scatterPlot}\ compares the \CI\ intensity with the \COt~\JJ{1}{0}\ and \JJ{2}{1}\ intensities \citep{Oka1998,Ginsburg2016}\ in voxel-by-voxel scatter plots, along with the best-fit correlation lines for the bulk component calculated with a robust fitting using the bi-weight algorithm.
The \CI\ intensity is tightly correlated with the \COt\ intensities in both plots. 
The \CI/\COt\ intensity ratios are 0.43 and 0.50 for the \COt\ \JJ{1}{0} and \JJ{2}{1}\ transitions, respectively.
Recent studies regarding the physical conditions of the CMZ consistently indicate the presence of at least two physical components:   the low-excitation component with gas kinetic temperature (\Tkin) of 20--50\ \kelvin\ and hydrogen volume density (\nHH) of $\sim 10^3\ \pcc$, and the high-excitation component with $\Tkin$ of $\sim100$ K and \nHH\ of $10^{4\text{--}5}\ \pcc$ \citep{Arai2016,Krieger2017,Mills2018,Tanaka2018b}.  
As the \COt~\JJ{1}{0}\ emission is dominated by the low-excitation component \citep{Tanaka2018b}, 
tight correlation of \CI\ with the low-$J$ \COt\ lines indicates that the \CI\ primarily originates from the low-excitation component.
In a later section (\S\ref{section:analysis:PCA}), 
we present a more systematic analysis to decompose the \CI\ distribution into contributions from the high- and low-excitation components 
using PCA.

\begin{figure*}[p]
\epsscale{1.0}
\plotone{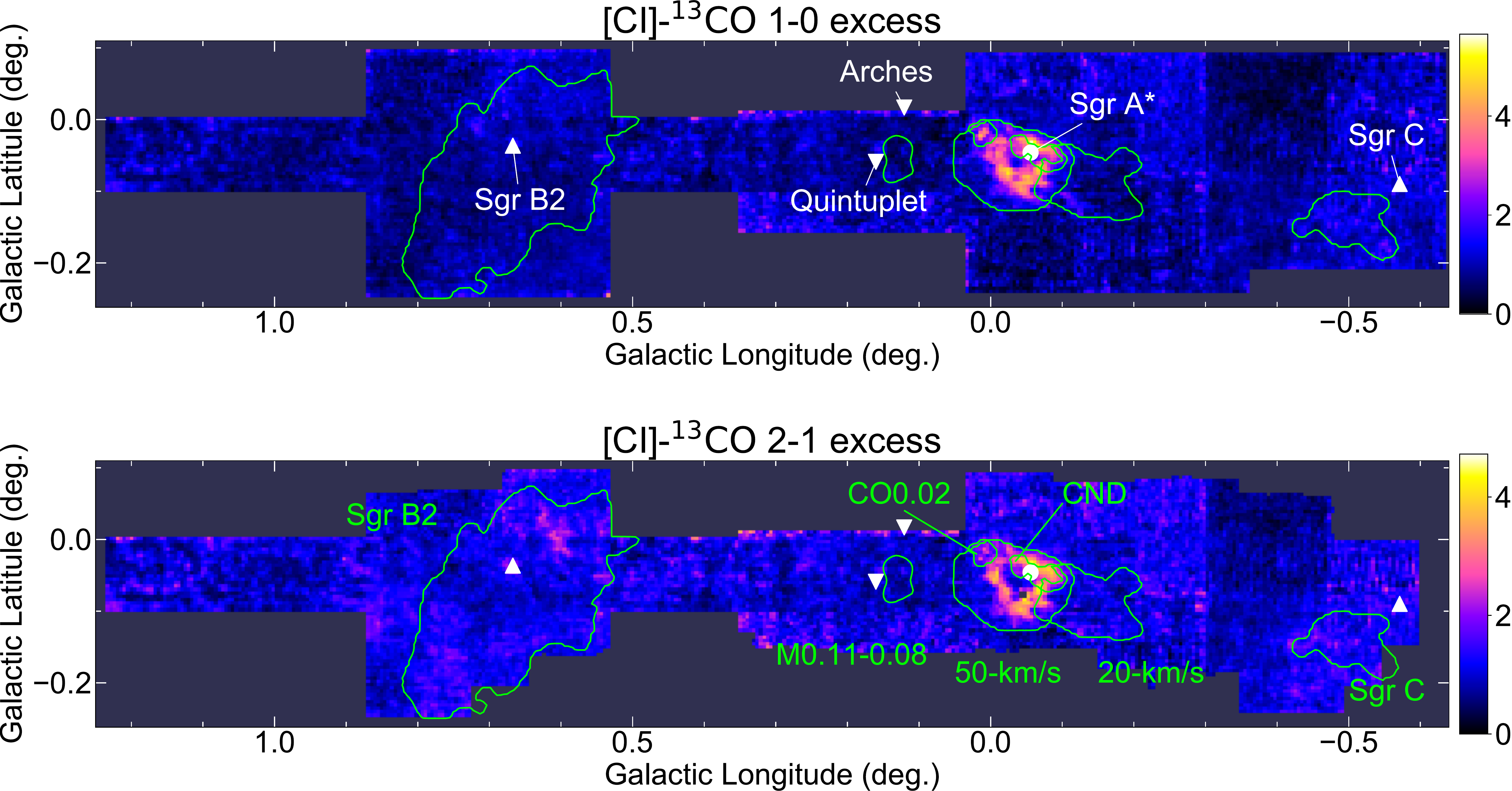}
\caption{Maps of the \CI\ excess from the overall \COt--\CI\ relation (Figure \ref{fig:13CO_CI_scatterPlot}),
  for the \CI-\COt\ \JJ{1}{0}\ (top) and the \CI-\COt\ \JJ{2}{1}\ (bottom), where the peak values in the 5-channel
  binned spectra toward individual line-of-sights are plotted.
  The positions of the nuclear
  blackhole (\sgras), young massive clusters (Quintupled and Arches cluster), and cluster-forming regions
  (Sgr B2 and Sgr C) are indicated in the figure.
  \label{fig:dCI}}

\plotone{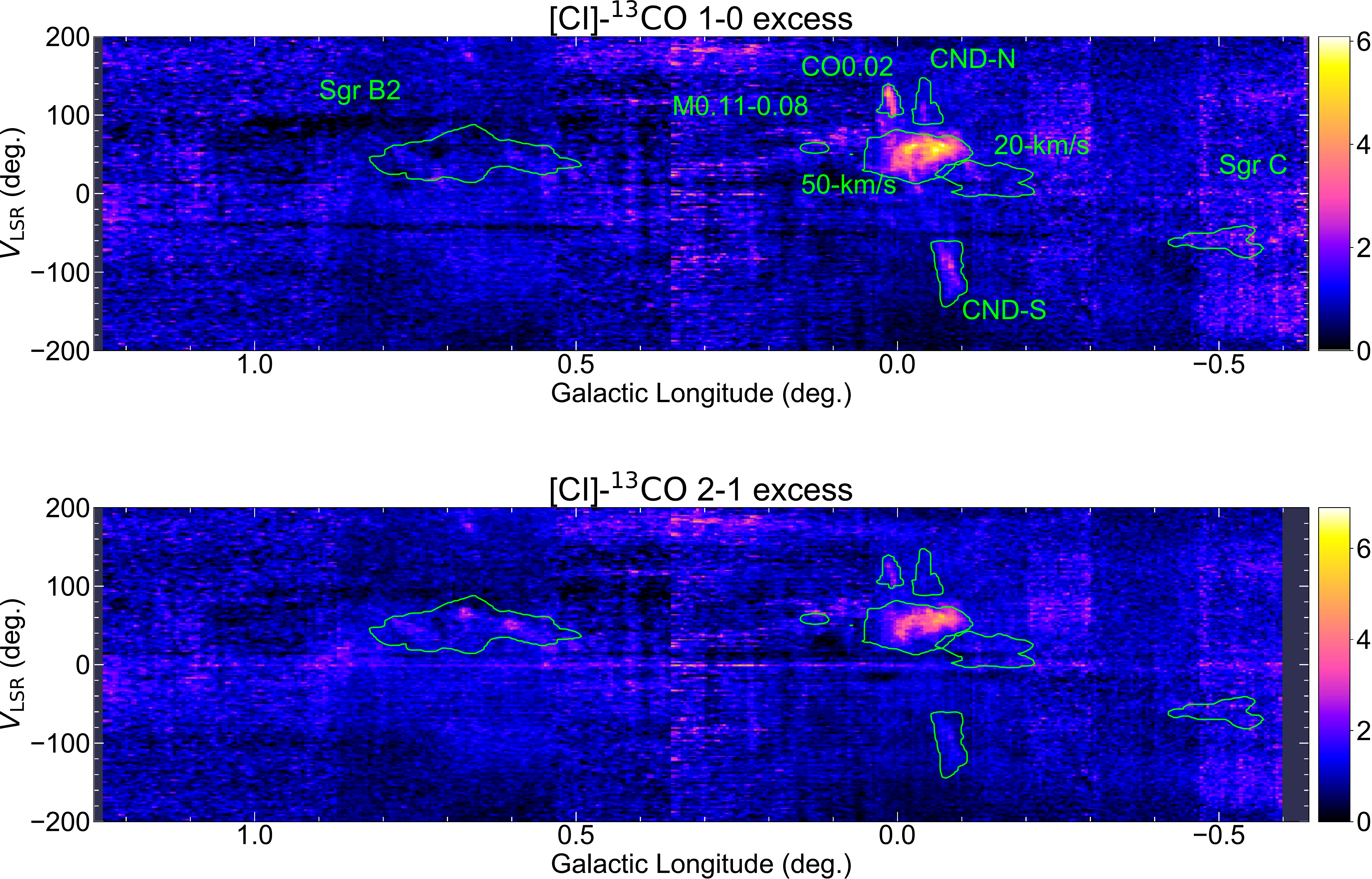}
\caption{Same as Figure \ref{fig:dCI}\ on the \gl--\vlsr\ plane.  
  \label{fig:dCIpv}}
  
\end{figure*}

The \COt--\CI\ scatter plots in Figure \ref{fig:13CO_CI_scatterPlot} show outliers located significantly above the best-fit correlation lines.
We calculate the excess \CI\ intensity from the \COt--\CI\ correlation, which we define as ${\dCI}_{1\text{--}0,\,2\text{--}1} \equiv \Tci - {\RCI}_{1\text{--}0,2\text{--}1}\cdot {\Tct}_{1\text{--}0,\,2\text{--}1}$, where \Tci\ and \Tct\ are the \CI\ \CIa\ and \COt\ intensities.
The subscripts $1\text{--}0$ and $2\text{--}1$ denote the \JJ{1}{0}\ and \JJ{2}{1}\ transitions of \COt, respectively.  
The factor \RCI\ is the overall \CI/\COt\ intensity ratio obtained with the robust fitting, namely, $\RCI_{1\text{--}0,\,2\text{--}1}$ = 0.43 and 0.50.
The position-position (PP) and  position-velocity (PV) distributions of the $\dCI_{1\text{--}0,2\text{--}1}$ are shown in Figures
\ref{fig:dCI} and \ref{fig:dCIpv}, respectively, in which we show the maximum \dCI\ along the axes perpendicular to the respective projection planes.
In both the $\dCI_{1\text{--}0}$ and $\dCI_{2\text{--}1}$ maps, 
voxels with large \dCI\ values appear predominantly in the Galactic-eastern half of the Sgr~A complex, 
containing the CND, the high-velocity portion ($\vlsr > 40\ \kmps$) of the 50-\kmps\ cloud, and CO0.02;
this \dCI\ distribution is almost identical to the distribution of the \CI-enhanced regions found in \cite{Tanaka2011}, despite the extended mapping area in this study.  
These results confirm that the enhanced \CI\ emission is exclusively present in the eastern part of the Sgr~A complex, and the \CI/\COt\ intensity ratio is approximately constant at larger radii.  

\subsection{[CI]-bright Ring Around the SNR Sgr~A~East}
\begin{figure*}
\begin{center}
    \epsscale{.65}
    \plotone{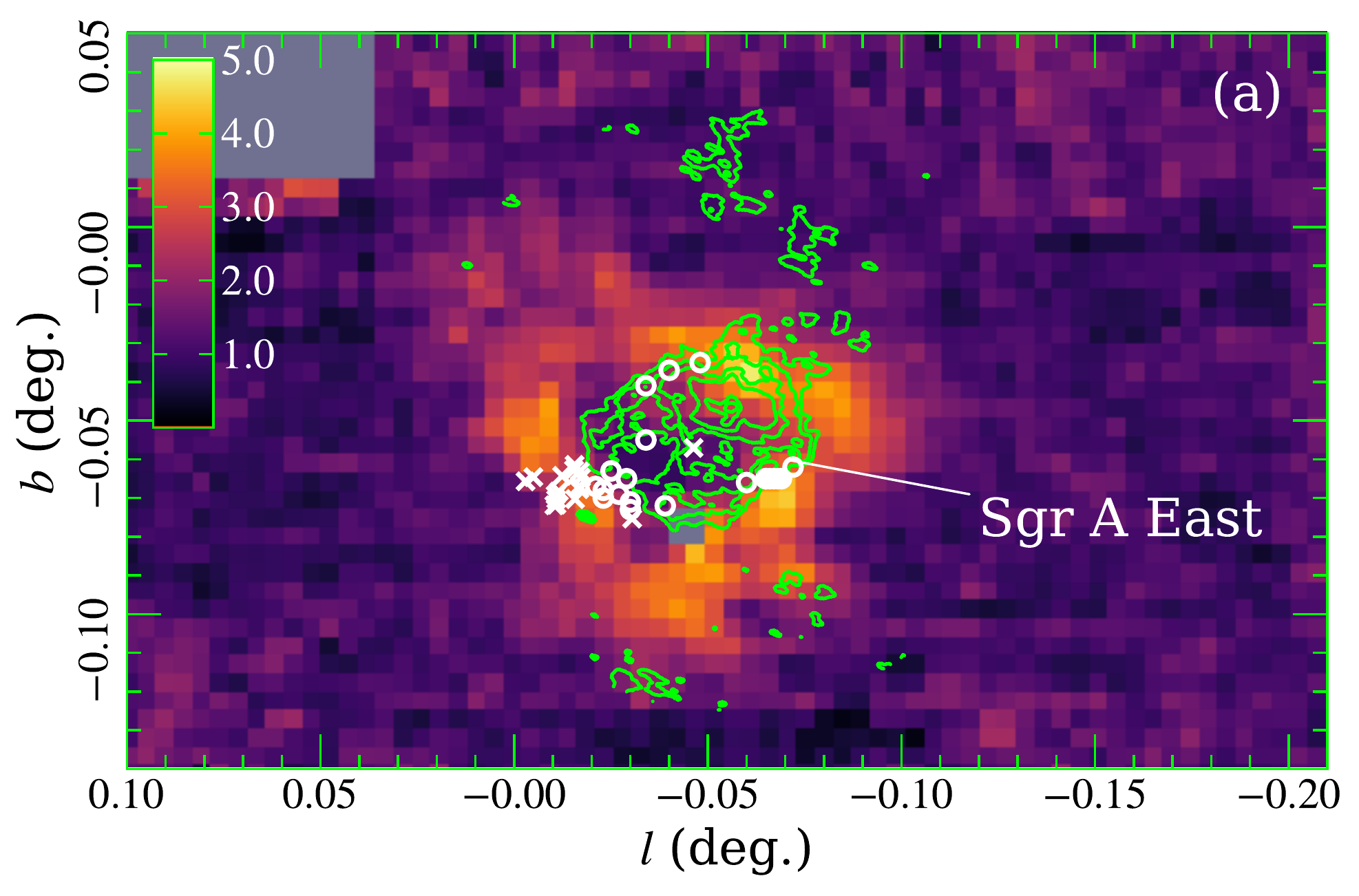}\\
    \plotone{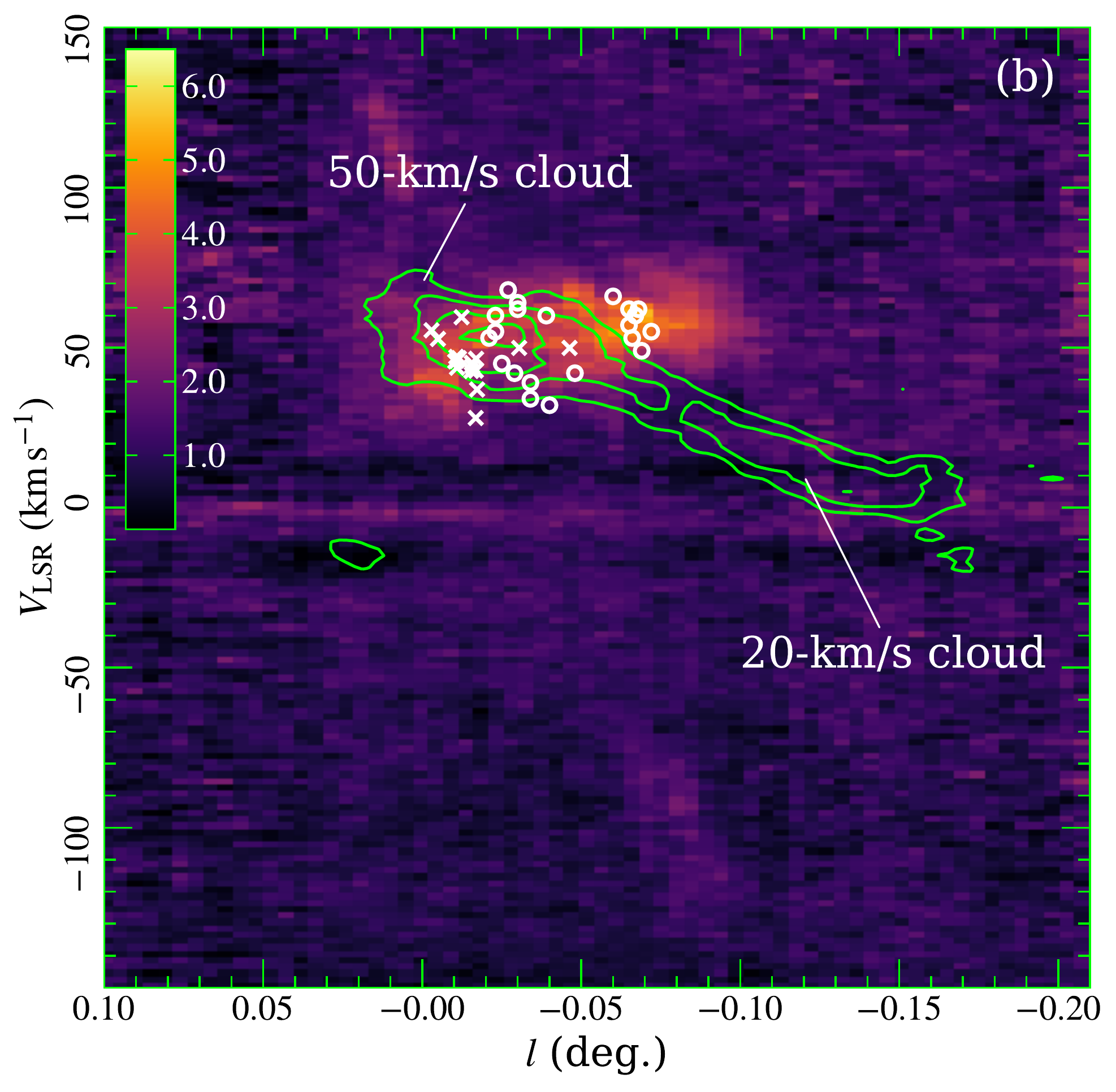}
\end{center}
\caption{ Close up views of the \CI--\COt\ \JJ{2}{1}\ excess image in the PP and PV plots (panels a and b, respectively).  Contours of the 6 cm continuum image taken from the VLA data archive (project code AZ0044) are overlaid on the top panel, at contour levels of 20, 50, 100, 200, 500, and 1000 ${\rm mJy}\,{\rm beam}^{-1}$.  The contour plot on the bottom panel is the \COt \JJ{3}{2}\ PV diagram averaged over the \gb\ range from $-0.085^\circ$ to $-0.061^\circ$, drawn at every 2 K interval starting from 5 K.  {The positions of the OH and class-I methanol masers \citep{Sjouwerman2008,Pihlstrom2011} are denoted by crossed and open circles, respectively.}
  \label{fig:closeUpCIcloud}}
\end{figure*}

Figure \ref{fig:closeUpCIcloud} shows close-up \dCIy\ images of the Sgr~A complex in the PP and PV views. 
The \CI-bright emissions exhibit a distinct ring-like structure unrecognized in the previous study {\citep{Tanaka2011}}.
The lower-left half of the ring corresponds to the curved ridge of the 50-\kmps\ cloud.
The upper-right half of the ring partly overlaps with the emission near the CND; however, this part is more spatially extended in the Galactic NE--SW direction while confined in a narrower velocity range (50--70\ \kmps) than the CND emission.
The \CI-bright ring as a whole encircles the outer edge of the radio shell of the SNR Sgr~A~East, indicating that the ring-shape formed via the interaction of Sgr~A~East with the surrounding GMCs  \citep[e.g.,][]{Ho1985,Sjouwerman2008,Tsuboi2011}.

Signatures of the interaction between the \CI-bright ring and Sgr~A~East can also be identified in the velocity structure of the \CI-excess emission.
Figure \ref{fig:closeUpCIcloud} compares the PV distributions of $\dCI_{2\text{--}1}$\ and \COt~\JJ{3}{2}.
The \COt~\JJ{3}{2}\ represents global kinematics of the Sgr~A complex characterized by the steep velocity gradient from the 20-\kmps\ cloud to the 50-\kmps\ cloud, which is interpreted as streaming or rotating motion {\citep{Sofue1995,Molinari2011,Kruijssen2015, Henshaw2016}. }
In contrast, \dCIy\ map does not show a noticeable velocity gradient; the \dCIy\ and \COt~\JJ{3}{2}\ peaks have approximately the same velocity in the eastern part of the ring, whereas their velocities deviate by 30\ \kmps\ at the western rim.    
This velocity structure of the \CI-bright ring is inconsistent with the global kinematics of the Sgr~A complex, likely as a consequence of the interaction with the Sgr~A East.
%\begin{revision}
The PPV distribution of the \dCIy\ is similar to that of the 1720 MHz OH masers and class-I methanol masers associated with the SNR shock \citep{Sjouwerman2008,Pihlstrom2011}. 
In particular, the SNR masers appears in the same velocity range as the \CI-bright ring (\vlsr = 50--70\ \kmps) in the western part of the SNR shell  ($\gl\sim -0.07^\circ$ to $-0.06^\circ$), where intense \COt\ emission is absent.  
%\end{revision}
%The 1720 MHz OH masers and class-I methanol masers detected along the SNR shell are distributed in the \vlsr\ range of 55--70 \kmps\ \citep{Sjouwerman2008,Pihlstrom2011}, being consistent with the \vlsr\ range of the \CI-bright ring.

In the Galactic disk region, enhancement of the \CI\ emission in SNR--molecular cloud interacting regions is reported in clump~C of IC443 \citep{White1994} and W51C \citep{Arikawa1999}, whereas no \Cn-enrichment was found in the molecular clouds near the SNR Cas~A \citep{Mookerjea2006}. 
The \CI-bright ring encompassing Sgr~A~East is the third example of the \CI-enhancement in a molecular cloud--SNR interaction system. 
This discovery suggests that the chemical processes associated with SN shocks or CRs  accelerated by the SNR are responsible for the \CI-enhancement in the Sgr~A complex \citep{White1994}.
We will present a detailed discussion regarding the origin of the \CI-bright ring in \S\ref{section:originOfCI}.

%Greatのあれとの比較
  
\subsection{Principal Component Analysis\label{section:analysis:PCA}}

\begin{deluxetable}{llccc}
\tabletypesize{\small}
\tablecaption{Lines Used in the Principal Component Analysis \label{table:PCAlines}}
%\documentclass[preprint]{aastex}

%\input{defines.tex}
%\usepackage{longtable}
%\usepackage{multirow}
%\begin{document}
%\begin{deluxetable}{lcccc}

%%%%%%%
\tablecolumns{5}
\small
\tablehead{\\ \colhead{Molecule} & \colhead{Transition} & \colhead{${\Eu}/{\kB}$$^{\rm a}$} & \colhead{${\rm log}_{10}{\ncrit}$$^{\rm a,b}$} & \colhead{Ref.$^{\rm c}$} \\
  & & \colhead{(K)} & \colhead{($\pcc$)} &  }
 \colnumbers
%\\
%\Colhead{8}{c}{(Jy\,\kmps)}}
\tabletypesize{\small}
\tablewidth{0pt}
\startdata
\CI               & \CIa                        & 23.6\phn & 3.0 & (1)\\
CO                & 1--0              & 5.3  & 3.3 & (2) \\
\COt              & 1--0              & 5.3  & 3.3 & (2) \\
                  & 2--1              & 15.9\phn & 3.8 & (3)\\
HCN               & 1--0              &  4.3 & 6.4 & (4)\\
                  & 4--3              & 42.5\phn & 7.3 & (5)      \\
\HCNt             & 1--0              & 4.1  & 6.5 & (5) \\
HNC               & 1--0              & 4.4  & 5.6 & (5)\\
\HCOp             & 1--0              &  4.3 & 5.3 & (4)\\
\HCOpt            & 1--0              &  4.2 & 5.3 & (6)\\
\NNHp             & 1--0              &  4.5 & 5.3 & (4)\\
CN                & $1_{3/2}$--$0_{1/2}$        & 5.4 & 4.4 & (1)\\
HNCO        & 4(0,4)--3(0,3)    & 10.5\phn & 5.0 & (4) \\
\HCCCN            & 10--9             & 24.0\phn & 5.2 & (5)\\
CS                & 1--0              &  7.1 & 5.3 & (7)\\
SiO               & 2--1              &  6.3 & 5.4 & (6)\\
${\rm C_2H}$        & $1_{1/2}$--$0_{1/2}$    &  4.2 & 4.5 & (4) 
\enddata
\tablenotetext{a}{Taken from the Leiden Molecular and Atomic Database \citep{Schoier2005}.}
\tablenotetext{b}{Values at $T = 50\ \kelvin$, calculated as $\ncrit = A_{i,i-1} / \mathop{\sum}\limits_{j <  i} C_{i,j}$, where $A$ and $C$ denote the Einstein A coefficient and collision rate coefficient, respectively. }
\tablenotetext{c}{References: (1) this work,  (2) \cite{Oka1998},  (3) \cite{Ginsburg2016}, (4) \cite{Jones2012}, (5) \cite{Tanaka2018b}, 
(6) \cite{Tsuboi2015}, (7) \cite{Tsuboi1999}}

%%%%%%%%%%
%\end{deluxetable}
%\end{document}

\end{deluxetable}

We apply PCA to multiline maps of the Sgr~A complex to
 characterize the physical and chemical properties of the \CI-enhanced regions.
PCA treats individual voxels as $N$-dimensional vectors, whose
elements are the intensities of the lines analyzed, with $N$ being the
number of lines.  The first principal component (PC1) is defined as
the direction of the maximum dispersion, and later PCs are
defined as the axes of the maximum dispersion in the subspace
that is perpendicular to all earlier PCs.
The directions of the PCs are given by the {eigenvectors} of 
the variance--covariance matrix of the original parameter space;
we can identify the lines with similar morphological characteristics according to the loci in the $N$-dimensional space of the  {eigenvectors}.
{Details of PCA are given in Appendix \ref{appendix:PCA}.}
%we refer the readers to, e.g., the application to the multiline data of the OMC-1 by 
%\cite{Ungerechts1997}.

\begin{figure*}
  \centering
  \epsscale{1.2}
    \plotone{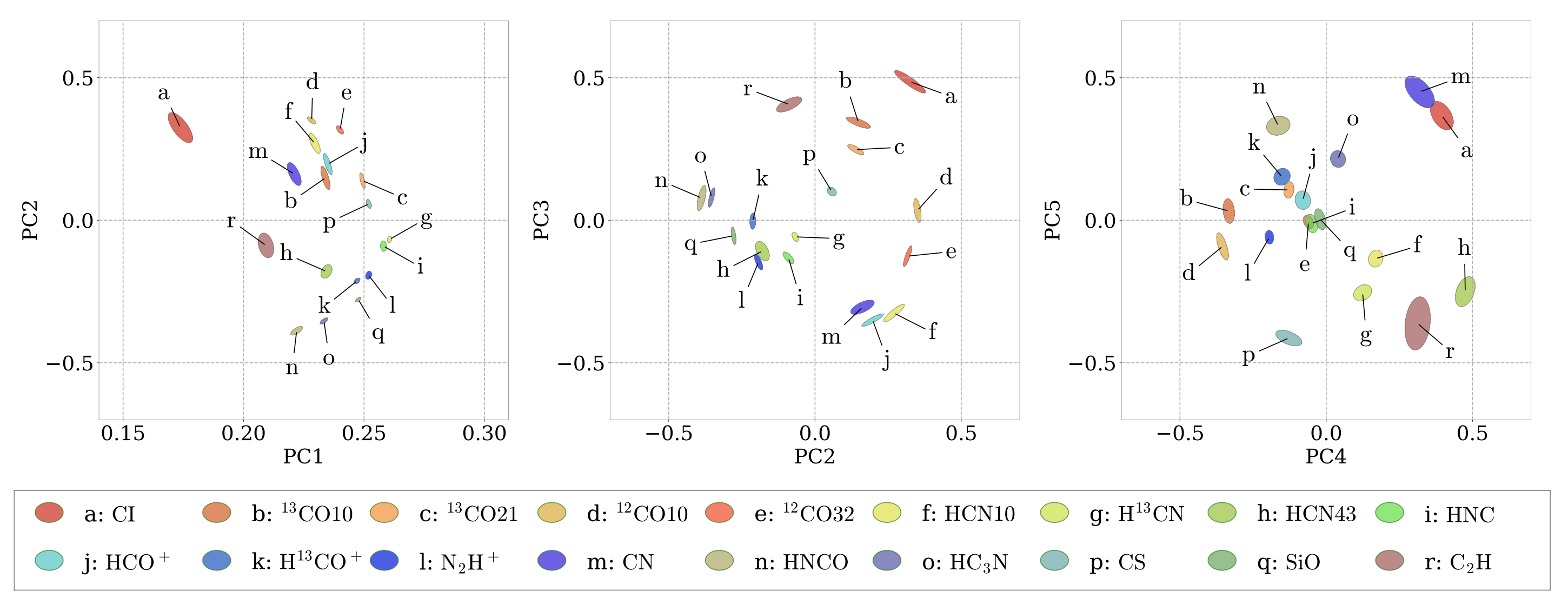}
    \caption{PC loading diagrams on the PC1--PC2, PC2--PC3, and PC4--PC5 spaces (from left to right).  The filled-ellipses indicate 3-$\sigma$ uncertainty estimated using the jackknife method.
    The signs of the vector elements are defined so that all 
    elements for \CI\ are positive.
    }
    \label{fig:PCloadingVectors}
\end{figure*}

\begin{figure*}
    \centering
    \epsscale{1.15}
    \plotone{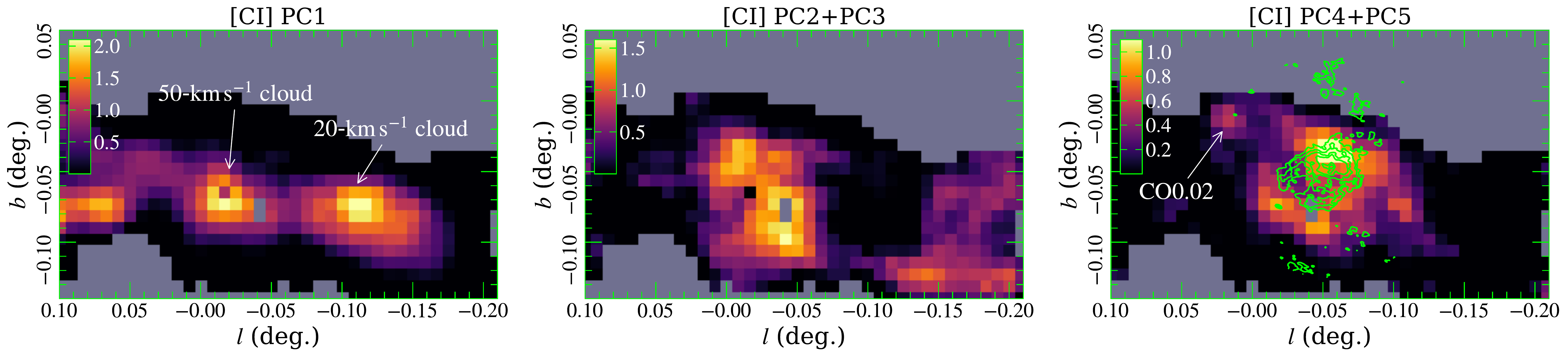}
    \caption{The \CI\ peak-intensity maps projected onto the PC1, PC2--PC3, and PC4--PC5 subspaces.  The overlaid contour plot is the 6 cm continuum emission same as that in Figure \ref{fig:closeUpCIcloud}.
    \label{fig:pc45score_synth}}
\end{figure*}

The 17 lines listed in Table \ref{table:PCAlines} are used for the analysis.
All data cubes are resampled into $30''\times30''\times5\,\kmps$ PPV bins with a Gaussian kernel of $60''$ FWHM and normalized so that the average and the standard deviation of individual lines are 0 and 1, respectively.
Voxel bins with  \CI~\CIa\ intensity below 1 K are removed. %to avoid disturbance from noises.  
Figure \ref{fig:PCloadingVectors} plots the {PC loading diagrams, which shows the eigenvectors on the PC1--PC2, PC2--PC3, and PC4--PC5 planes}, along with their 3-$\sigma$ uncertainties estimated using the jackknife method.
The cumulative contribution ratio up to PC5 is 92.8\%, indicating that the distributions of the molecular lines involved in the analysis can be decomposed into the five characteristic components with sufficient accuracy.
Figure \ref{fig:pc45score_synth}\ shows the PP distribution of \CI\ in the Sgr~A complex, projected onto the subspaces of PC1, PC2--PC3, and PC4--PC5.

In the PC1--PC2 and PC2--PC3 loading diagrams,
\CI\ (labeled a in the plot) is at a position close to \COt~\JJ{1}{0}\ and ~\JJ{2}{1}\ (labeled b and c, respectively), except for that \CI\ has a slightly smaller PC1 loading and slightly greater PC2 and PC3 loadings than the \COt\ lines.   
Therefore, PCs 1--3 can be regarded representing the bulk component with uniform \CI/\COt\ intensity ratios identified in the scatter plots (Figure  \ref{fig:13CO_CI_scatterPlot}).
The morphological difference between the \CI\ and \COt\ lines  appears in the PC4--PC5 diagram;  \CI\ has large positive PC4 and PC5 loadings, whereas the \COt\ lines show negative PC4 loadings and near-zero PC5 loadings.  
The PC4+5 component of the \CI\ map (Figure \ref{fig:pc45score_synth}) exhibits the characteristic ring-shape that is almost identical to that of the \dCI\ maps (Figure \ref{fig:closeUpCIcloud}).  These results clearly indicate that  the \CI-enhanced region in the Sgr~A complex is represented by PC4 and PC5 without significant contributions from other PCs.

The PCA results reveal a few important chemical and physical characteristics of the \CI-emitting region.   First,
the PC4--PC5 loading diagram indicates that the CN intensity (labeled m in the figure) also increases in the \CI-enhanced region; \CI\ and CN show large positive PC4 and PC5 loadings, being clearly isolated from all other lines in the diagram.
It is noteworthy that neither optically thin shock tracers (SiO and \HCCCN) nor quiescent-gas tracers (\NNHp, \HCOpt, HNC) are clustered in the PC4--PC5 loading diagram.
The previous analysis using the entire CMZ data without \CI\ and CN \citep{Tanaka2018b} showed that the clusters of shock and quiescent-gas tracers are relatively well separated from each other in the loading diagram, indicating that shock chemistry is one of leading factors in the molecular cloud chemistry in the CMZ. 
Therefore, the absence of the cluster of shock/quiescent-gas tracers in Figure \ref{fig:PCloadingVectors}\ indicates that the enhancement of \CI\ and CN is likely caused by a factor other than shock chemistry. 

We also find that the optically thin low-density tracers (\CI\ and \COt) and the optically thin high-density tracers (HNC, \HCNt, \HCOpt, \NNHp) are loosely clustered in the PC2--PC3 loading diagram.  
As the \COt\ lines and the dense-gas tracers primarily trace the low- and high-excitation components of the CMZ clouds \citep{Mills2018,Tanaka2018b}, respectively, this result indicates that the distributions of the low and high excitation components are clearly different from each other.
The PC2+PC3 component map (Figure \ref{fig:pc45score_synth}) shows that that the low-excitation component has more spatially extended distribution than the PC1 map, as reasonably expected. 
\CI\ has the largest PC2 and PC3 loadings and the smallest PC1 loading of all lines analyzed, suggesting that \CI\ predominantly originates from the low-excitation component.
If we decompose the \CI\ distribution into two components proportional to the \COt~\JJ{1}{0}\ and HNC~\JJ{1}{0}\ distributions in the PC1--3 space, linear regression analysis gives $S_{\rm CI} = (1.409\pm0.004)\cdot S_{\COt} - (0.588\pm0.004)\cdot S_{\rm HNC}$, where vector $S_X$ is the normalized intensity of line $X$; 
the negative coefficient for $S_{\rm HNC}$ could be interpreted as  that \CI\ traces the low-excitation component better than \COt~\JJ{1}{0}.

% some nice analysis

\subsection{Abundance Ratios\label{section:analysis:excitationAnalysis}}

\subsubsection{{\NN(\Cn)/\NN(\CO)} Ratio}

We evaluate the \Xcn\ abundance ratio of the entire CMZ and the 7 regions listed in Table \ref{table:CIabundance}.
The boundaries of the selected regions are denoted by contours in the \gl--\gb\ and \gb--\vlsr\ spaces in Figures \ref{fig:ppmap}\ and \ref{fig:dCIpv}.
The input parameters are the averaged intensity ratios among \COt\ \JJ{1}{0}, \JJ{2}{1}, CO \JJ{1}{0}, and \CI, whose values are listed in Table \ref{table:CIabundance}.
The Bayesian framework is used to calculate the credible interval of the \Xcn\ abundance ratio, assuming log-uniform prior distributions of the CO column density, \nHH, and \Tkin\ in the finite ranges of $10^{20}$--$10^{22}$\ $\psc$, $10^2$--$10^{4}$\ $\pcc$, and 20--50 K, respectively, which are sufficiently wide to cover the physical conditions in the low-excitation component \citep{Nagai2007,Krieger2017,Mills2018,Tanaka2018b}.
The {\NN(\CO)/\NN(\COt)} abundance ratio is fixed at 24 \citep{Langer1990,Langer1993,Tanaka2018b}.
Line intensities are calculated using the rate coefficients taken from the Leiden Molecular and Atomic database (LAMDA; \citealt{Schoier2005}) on the large-velocity gradient (LVG) approximation \citep{Goldreich1974}.
We adopt a simple one-zone approximation by ignoring the contribution from the high-excitation component, because the \CI\ and low-$J$ CO and \COt\ lines predominantly originate from the low-excitation component, as seen in the previous section (\S\ref{section:analysis:PCA}).

\begin{figure*}
    \epsscale{1.15}
    \plotone{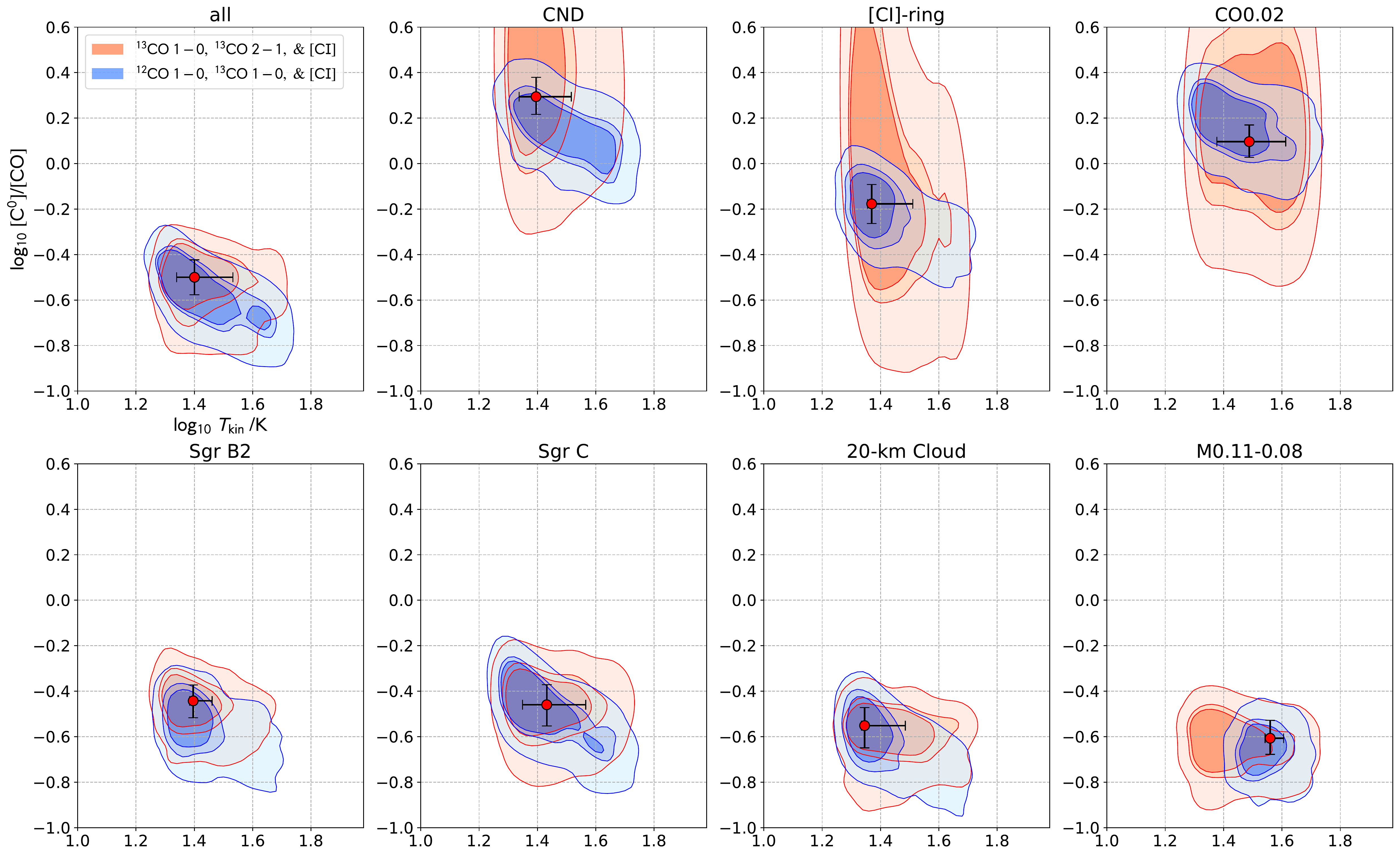}
    \caption{Simultaneous credible intervals of \Xcn\ and \Tkin\ for individual regions.   The orange and blue contours are the credible intervals calculated using the data sets of \{\CI, \COt~\JJ{1}{0}, \COt~\JJ{2}{1}\} and \{\CI, \COt~\JJ{1}{0}, \CO~\JJ{1}{0}\}, respectively.  The contours are drawn at 50, 65, and 95\%\ credible intervals.  The median values using all data are denoted by red filled circles with 1-$\sigma$ bars. }
    \label{fig:LVGres2}
\end{figure*}

Figure \ref{fig:LVGres2}\ shows the simultaneous credible intervals of the {\NN(\Cn)/\NN(\CO)} abundance ratio and \Tkin.
We confirm that the variation of the \CI/\COt\ intensity ratio mainly reflects that in the  {\NN(\Cn)/\NN(\CO)} abundance ratio, and that the variation in the excitation condition among the regions is negligible.
The median \Tkin\ values are in a narrow range of 20--35 K for all regions.
{The CMZ clouds without \CI-enhancement} (Sgr~B2, Sgr~C, 20-\kmps~cloud, and M0.11{$-0.08$}) have approximately uniform {\NN(\Cn)/\NN(\CO)} abundance ratios of 0.2--0.4.
The \Cn~abundances in the \CI-bright clouds (CND, \CI-bright ring, and CO0.02) are significantly enhanced by a factor of $\gtrsim 2$ compared with the {clouds without \CI-enhancement}.  

This analysis confirms that the CND and CO0.02 have remarkably 
high \Xcn\ ratios of $> 1$, although it was unknown whether their high \CI/\COt\ intensity ratios translate into enhanced \Cn\ abundances or high \COt\ excitation temperatures in the previous paper \citep{Tanaka2011}.
Now, we find that the \COt~\JJ{2}{1}\ to \JJ{1}{0}\ ratios in these regions do not differ much from the CMZ average, indicating highly enhanced \Cn\ abundances.
In particular, our non-LTE analysis has shown \Tkin = 20--30 K in the CND, apparently contradictory to the much higher \Tkin\ of $> 200\ \kelvin$ calculated with CO SED analysis \citep{Requena-Torres2012}. 
The reason for this inconsistency is that the analysis in \cite{Requena-Torres2012} uses high-$J$ CO levels up to $J$=16 and hence is biased to the high-excitation component invisible in the low-$J$ lines.
Our analysis indicates that the low-$J$ \COt\ emission is dominated by the low-excitation component even in the circumnuclear region.
This low-excitation gas could be the molecular gas counterpart of the cold dust component with the 23.5~K dust temperature \citep{Etxaluze2011}.

\begin{deluxetable*}{lccccccccc}
\tablecaption{Averaged Line Intensities and abundance ratios\label{table:CIabundance}}
\tablehead{ & \multicolumn{6}{c}{$\Tmb$~/K} & \colhead{{\NN(\Cn)/\NN(\CO)}} & \multicolumn{2}{c}{{\NN(CN)/\NN(HCN)}} \\
\cline{2-7}\cline{9-10}
\colhead{} &\colhead{\CI} & \colhead{\COt~1\text{--}0} & \colhead{\COt~2\text{--}1} & \colhead{CO~1\text{--}0}  & \colhead{CN} &\colhead{\HCNt~1\text{--}0} &
& \colhead{low-ex.} & \colhead{high-ex.}}
\colnumbers
%\\
%\Colhead{8}{c}{(Jy\,\kmps)}}
\tabletypesize{\small}
\tablewidth{0pt}
\startdata
All & $1.6$ & $3.4$ & $2.7$ & $13$ & \nodata & \nodata & $0.32^{+0.06}_{-0.05}$ & \nodata \\
\multirow{3}{*}{CND}  & \multirow{3}{*}{$0.73$} & \multirow{3}{*}{$0.44$} & \multirow{3}{*}{$0.31$} & \multirow{3}{*}{$2.2$} & \multirow{3}{*}{$0.21$} & \multirow{3}{*}{$0.045$} &
\multirow{3}{*}{$1.95^{+0.41}_{-0.31}$} & \multirow{3}{*}{$0.24$} & $ 0.38\tablenotemark{a}$ \\ 
&&&&&&&&& $0.57\tablenotemark{b}$ \\
&&&&&&&&& $0.78\tablenotemark{c}$ \\
\CI-bright ring & $4.6$ & $4.9$ & $4.9$ & $18$ & $0.62$ & $0.51$  & $0.66^{+0.14}_{-0.12}$ &    $0.06$ & $ 0.10 $ \\
CO0.02 & $1.3$ & $0.68$ & $1.1$ & $7.8$ & $0.23$ & $0.14$ & $1.24^{+0.22}_{-0.19}$ & $0.07$ & $ 0.13 $ \\
Sgr~B2 & $3.0$ & $5.8$ & $4.1$ & $19$ & \nodata & \nodata & $0.36^{+0.06}_{-0.06}$ & \nodata  & \nodata \\
Sgr~C & $2.5$ & $4.3$ & $3.9$ & $14$ & \nodata & \nodata & $0.30^{+0.07}_{-0.07}$ & \nodata & \nodata \\
20-\kmps~Cloud & $2.8$ & $6.2$ & $5.5$ & $17$ & $0.60$ & $0.80$ & $0.27^{+0.06}_{-0.05}$ & $ 0.03 $ & $0.06$ \\
M$0.11$$-0.08$ & $3.4$ & $7.4$ & $6.7$ & $27$ & $0.68$ & $0.70$ & $0.25^{+0.05}_{-0.04}$ & $ 0.04 $ & $0.08$ \\ 
High-excitation &\nodata&\nodata&\nodata&\nodata& \nodata & \nodata& $\lesssim0.1$ & \nodata  & \nodata
\enddata
%\tablenotetext{a}{The CN intensities are summed over the five hyperfine components.}
\tablenotetext{a}{Assuming (\Tkin, \nHH) = $(80\ \kelvin, 10^{4.1}\ \pcc)$ \citep{Tanaka2018b}}
\tablenotetext{b}{Assuming (\Tkin, \nHH) = $(200\ \kelvin, 10^{4.5}\ \pcc)$ \citep{Requena-Torres2012}}
\tablenotetext{c}{Assuming (\Tkin, \nHH) = $(500\ \kelvin, 10^{5.2}\ \pcc)$ \citep{Requena-Torres2012}}

\end{deluxetable*}

\newcommand\Xhigh[1]{\ifmmode{X_\mathrm{high}\left({#1}\right)}\else{$X_\mathrm{high}\left({#1}\right)$}\fi}
\newcommand\RN{\ifmmode{R_N}\else{$R_N$}\fi}

The \Xcn\ abundance ratio in the high-excitation component is not directly known from the above excitation analysis; however, the PCA results indicate that the \Xcn\ ratio is likely suppressed in the high-excitation component.
As we found in \S\ref{section:analysis:PCA}, the fractional contribution from the high-excitation component in the total \CI\ intensity is less than that in \COt~\JJ{1}{0}.
Meanwhile, the \CI~\CIa~to~\COt~\JJ{1}{0}\ intensity ratio should increase with increasing \Tkin\ and \nHH;  
non-LTE excitation analysis shows that the intensity ratio in the high-excitation component 
is 2--3 times that in the low-excitation component if the \Cn\ and \COt\ abundances are constant across the two components, with $(\Tkin, \nHH) = (80\ \kelvin, 10^{4.1}\ \pcc)$ and $(20\text{--}40\ \kelvin, 10^{3.0\text{--}3.2}\ \pcc)$ being assumed for the high- and low-excitation components, respectively \citep{Tanaka2018b}.
Therefore, the \Cn\ abundance in the high-excitation component should be suppressed at least by a factor of 2--3, yielding $\Xcn\lesssim 0.1$.

\subsubsection{{\NN(\CN)/\NN(HCN)} Ratio}
The CN abundance is difficult to accurately calculate mainly due to the limited information of the physical conditions of the CN-emitting region.
The locus of the CN in the PC2--PC3 loading diagram (Figure \ref{fig:PCloadingVectors}) is closer to high-density tracers such as \HCNt\  and \HCOpt, than to low-density tracers such as \COt\ and \CI;  however, the closeness of CN to \CI\ in the PC4--PC5 loading diagram suggests that CN emission may contain a contribution from the low-excitation component traced by \CI\ at least in the \CI-enhanced regions.  
Therefore, we calculate {\NN(CN)/\NN(HCN)} abundance ratios both for the low- and high-excitation cases.   The physical condition parameters of \Tkin, \nHH = 20 K, $10^{3}\ \pcc$ and 80 K, $10^{4.1}\ \pcc$ are used for the low- and high-excitation cases, respectively.
For the CND, we also show the abundance ratios calculated using the parameters in the two-component model by \cite{Requena-Torres2012} in the table.
The excitation calculation is performed using the rate coefficients taken from the LAMDA on the optically-thin approximation. 
The results are tabulated in columns 9 and 10 of Table \ref{table:CIabundance}.

In either of the low- and high-excitation cases, 
the {\NN(CN)/\NN(HCN)} abundance ratios in 
the \CI-bright clouds (CND, \CI-bright ring, and CO0.02) are a factor of $\sim$ 1.5--2 higher than in {those without \CI-enhancement }(the 20-\kmps\ cloud and M0.011$-0.08$).   The CND has a remarkably high CN abundance, whose {\NN(CN)/\NN(HCN)} ratio is 5 times that of {the clouds without \CI-enhancement} even with the conservative assumption that \Tkin\ and \nHH\ of the CND are the same as the CMZ averages. 
The physical condition parameters from \cite{Requena-Torres2012} yield higher {\NN(CN)/\NN(HCN)} ratios of 0.57--0.78, being consistent with the abundance ratio measured by \cite{Harada2015}.
We note that our results likely underestimate the {\NN(CN)/\NN(HCN)} abundance ratio if CN is not optically thin. 
\cite{Martin2012} measured the CN opacity to be 2--3 toward bright CN peaks in the CND;
if the entire CND has this high CN opacity, the opacity corrected CN abundance is 2--3 times the values in Table \ref{table:CIabundance}.

\section{DISCUSSION\label{section:discussion}}

\subsection{Possible Mechanisms of the \Cn-Enrichment \label{section:originOfCI}}
In this subsection, we discuss the possible origins of the \Cn-rich state in the CMZ clouds by mainly focusing on the \CI-bright ring.
The spatial correlation between the \CI-bright ring and the radio shell of the Sgr~A~east SNR 
suggests that the SNR--molecular cloud interacting regions plays a dominant role in the \Cn-enrichment, as reported in 
SNR--GMC interacting system in the Galactic disk region \citep{White1994,Arikawa1999}.
The analysis described in \S\ref{section:analysis:PCA} and \S\ref{section:analysis:excitationAnalysis} has shown that the enhancement of the \CI\ emission is caused by the \Cn-enrichment in the low-excitation component of the CMZ gas, which is accompanied by an enhancement of CN.
In the previous paper \citep{Tanaka2007}, we proposed four possible origins of the \CI-bright clouds, namely, shock dissociation, CR dissociation,  X-ray dissociation, and time-dependent PDR chemistry.
In the following subsections, we reexamine these hypotheses based on the above morphological and chemical characteristics of the \CI-bright ring.

\subsubsection{Shock Chemistry}
The SNR--GMC interaction may increase the \Cn\ abundance through two chemical processes: shock chemistry and CR induced chemistry.
The shock chemistry is less preferred of the two mechanisms,
because the simultaneous enhancement of \Cn and CN found with the PCA (\S\ref{section:analysis:PCA}) is atypical for shocked molecular gas.
Observations toward the interacting SNR IC443 \citep{Ziurys1989,Turner1989} and outflow sources in star-forming regions \citep{Blake1987,Rodriguez-Fernandez2010,Shimajiri2017} found that CN mainly traces quiescent ambient gas whereas no particular increase of CN is detected in shocked regions.
The inability of the CN to trace shocked gas is consistent with 
the model by \cite{Mitchell1984}, in which the CN decreases in the shock velocity regime where the {\NN(\Cn)/\NN(\CO)} abundance ratio is $>1$.
The \CI-bright ring also differs from typical SN-shocked gas in the absence of enhancement in the shock tracer lines such as SiO;
conversely, many of the shocked molecular clouds in the CMZ with enhanced SiO emission or broad velocity wings  \citep{Tsuboi2015,Tanaka2014,Tanaka2015,Tanaka2018a} does not show enhanced \CI\ emission in our data.

\subsubsection{Cosmic-ray-induced chemistry}

\newcommand{\NII}{\ion{N}{2}}
The low energy ($\lesssim $ GeV) CRs created in SNR-shocked interstellar medium efficiently increase the \Cn\ abundance through dissociation of CO by CR particles or by CR-ionized He \citep{Boger2005,Papadopoulos2010,Bisbas2015,Bisbas2017,Papadopoulos2018}.
An advantage of this CR dominated region (CRDR) picture over the shock chemistry is that it more easily {explains} the observed enhanced \Cn\ and CN abundances.
\cite{Boger2005} shows that the \Cn\ and CN abundances almost monotonically increase with increasing $\zeta\nHH^{-1}$. 
So does the {\NN(CN)/\NN(HCN)} abundance ratio, except for a small discontinuous drop at the transition from the low ionization phase from the high ionization phase.    The observed {\NN(\Cn)/\NN(CO)} abundance ratio of 0.7 and {\NN(CN)/\NN(HCN)} abundance ratio of 0.1 approximately correspond to $\zeta \sim 10^{-16}\cdot\left(\frac{\nHH}{10^3\ \pcc}\right) {\rm s}^{-1}$ in their calculation; i.e., an order enhancement of $\zeta$ from the canonical value ($\sim 10^{-17}\ \mathrm{s}^{-1}$) can explain the observed \Cn\ and CN abundances when $\nHH \sim 10^3\ \pcc$.

Observational studies consistently indicate elevated \zCR\ up to $10^{-14}\ \ps$ in the CMZ \citep{Oka2005,Goto2008,Indriolo2015,Petit2016,Takeshi2019,Willis2020}.  However, they are mostly based on the absorption study of \HHHp\ and $\mathrm{H}_{x}\mathrm{O^+}$ against bright sources, which is sensitive to \zCR\ in diffuse clouds with $\nHH\sim 10^2\ \pcc$;  the \zCR\ values in the interior of denser molecular gas visible in CO and \CI\ emissions are often substantially lower than those in diffuse clouds and molecular cloud surfaces \citep{Indriolo2007,Rimmer2012,Albertsson2018,Willis2020}.
No direct measurement of \zCR\ for the interior of the Sgr~A complex is present; however, we may refer to the measurements toward interacting SNRs of similar ages as Sgr~A~East in the Galactic disk region by  \cite{Ceccarelli2011} and \cite{Vaupre2014}.
They used the millimeter  \HCOp\ and $\mathrm{DCO^{+}}$\ emission lines to measure \zCR\ in dense molecular gas ($\nHH > 10^3\ \pcc$) in IC443 and W28, and obtained $\zCR\sim 10^{-15}\ \ps$.   If the same \zCR\ value applies to the Sgr~A~East, it is sufficiently high to create the \Cn-rich chemical composition.

In this CRDR picture, the \CI-bright ring approximately represents the distribution of the ionizing CRs that escaped from the acceleration site.
The outermost extension of the \CI-bright ring is $\sim 5\,\pc$ ahead of the radio synchrotron front of the SNR.
This is consistent with the spatial variation of \zCR\ in W28, in which enhanced \zCR\ values of $\sim 10^{-15}\ \ps$ are found at least $\sim 3$ pc away from the radio shell.   As W28 and Sgr~A~East are of similar ages ($1\text{--}10\times 10^4$ yr; \citealt{Maeda2002,Vaupre2014}), we may expect a similar spatial extension of ionizing CR in them, though the high magnetic field strength $B\sim 100\ \mu\mathrm{G}$ in the CMZ \citep{Crocker2010}\ may somewhat prevent the CR diffusion for the case of  Sgr~A~East.
By using the diffusion coefficient of CR particles of energy $E$, $D(E) = 10^{28}\left(\frac{E}{10\, \mathrm{GeV}}\right)^{0.5}\left(\frac{B}{3\,\mu\mathrm{G}}\right)^{-0.5}\,\mathrm{cm}^2\ps$ \citep{Gabici2009}, the diffusion distance at time $t$ is $\sqrt{4D(E)t} = (5\text{--}9)\left(\frac{t}{10^4\ \yr}\right)^{\frac{1}{2}}\ \pc$ for the energy $E$ of 0.1--1 GeV.
This diffusion distance is still sufficiently large to encompass the entire \CI-bright ring.

However, the CRDR picture does not explain all observed features. 
A problem with the CRDR picture is the absence of enhancement of the \NNHp\ and \HCOp\ abundance in the \CI-enhanced regions. 
In the PC4--PC5 loading diagram (Figure \ref{fig:PCloadingVectors}), 
these molecular ions are located close to the coordinate origin, indicative that their intensities do not differ between inside and outside the \CI-enhanced regions;
on the other hand, models predict increases in their abundances in high \zCR\ environments (e.g., \citealt{Caselli1998,Papadopoulos2007,Harada2015,Albertsson2018}).
This mismatch with the models remains as a problem to be solved. 
One possible explanation is that the \HCOpt\ and \NNHp\ emissions in the low-excitation component are not detectable due to their high \ncrit, which are approximately 1--2 orders of magnitude higher than those of \CI, \COt, and CN. 
Typical parameters of the low-excitation component (\Tkin, \nHH = 20 K, $10^3\ \pcc$) and \NNHp\ and \HCOp\ column densities of $10^{13}\ \psc$ per unit velocity width \citep{Tanaka2018b} yield the \HCOpt\ and \NNHp~\JJ{1}{0}\ intensities of 0.02 K and 0.01 K, respectively, which are below the noise levels of the data used in the PCA, $\sim 0.05$ K.

\subsubsection{XDR}
X-ray dissociation may create a chemical composition with rich \Cn\ and CN abundances in the vicinity of strong X-ray sources such as AGNs \citep{Meijerink2005,Meijerink2007,Izumi2020}. 
Past quasar activities of \sgras\ are suggested by the Fermi Bubbles \citep{Su2010} and the recently discovered 430-pc bipolar radio bubbles \citep{Heywood2019}; however, they are unlikely the immediate origin of the \CI-bright ring associated with the Sgr~A~East SNR, as the estimated ages of the bubbles of $1\ \Myr$ are substantially longer than the age of the Sgr~A~east SNR \citep[$\lesssim 10^4$ yers;][]{Maeda2002,Sakano2004}.
We may alternatively consider the short outburst of \sgras\ at a few 100 years ago suggested by the X-ray reflection nebulae {\citep{Koyama1996,Ponti2010,Clavel2013,Ryu2012}}, during which the X-ray luminosity ($L_X$) of \sgras\ was $10^6$ times the present value,  $10^{33}\ \erg\,\mathrm{s}^{-1}$.
However, it is questionable whether the \Cn\ abundance detectably increases during the short outburst.
The CO dissociation time scale is estimated to be a few $10^3\ \yr$ at the position of the \CI-ring when $L_X$ of \sgras\ is $10^{39}\ \erg\,\mathrm{s}^{-1}$ \citep{Maloney1996}; if we reasonably assume the duration time of the \sgras\ burst to be $\leq 10^2\ \yr$, it is more than an order of magnitude shorter than the time scale necessary for significant \Cn\ enhancement.

\subsubsection{Time-dependent chemistry}

%Photodissociation region (PDR) is an important origin of the interstellar \Cn\ and CN.
{Ultra violet photodissociation is one of the major formation processes of the interstellar \Cn\ and CN.}
The bright \CII, \NII, and 8 $\mu$m PAH emissions \citep{Stolovy2006,Garcia2016}\ indicate the formation of PDR in the Sgr A complex, which is exposed to intense UV radiation mainly from the central cluster.
However, UV photodissociation is not an efficient process to increase the \Cn\ abundance, as the \Cn\ in the PDR is confined within a thin layer regardless of the UV intensity \citep{Hollenbach1991,Meijerink2007}.
{Indeed, the southeastern portion of the \CI-bright ring is coextensive with the dense gas ridge of the 50-\kmps\ cloud although it is irradiated from the northwestern direction by the UV source.   This \CI\ distribution indicates that the majority of \Cn\ resides in the cloud interior, contradictory to the thin \CI\ layer predicted by the standard PDR model.  The co-extensive \CI\ and \COt\ emissions are also generally observed in the solar-neighborhood region with and without UV sources} \citep[e.g.,][]{Plume1994,Ikeda1999,Maezawa1999,Shimajiri2013}.
Moreover, the vast majority of other PDRs under the extreme UV flux from the Sgr~B2 proto-cluster and the Quintuplet cluster do not show \CI-enhancement, 
indicating that intense UV radiation field is unlikely the immediate origin of the \CI-enhancement.

However, the PDR picture may explain the \CI-bright ring shell if the time-dependent chemistry is considered.  
The \CI-bright ring should contain a significant amount of low-density material collected by the expanding SNR shell.
By using the formula by \cite{Papadopoulos2004}, the \Cn\ to CO conversion time scale in the medium of $\nHH\sim 10^3\ \pcc$ is $0.8~\Myr$, which is nearly two orders of magnitude longer than the age of the SNR Sgr~A~east, a few $(1\text{--}10)\times10^4\ \yr$; hence, the collected material is in the \Cn-dominant phase if the elemental carbon was in the form of $\Cn$ or $\Cp$ in the initial condition.
The CN abundance also increases in early phases in the time-dependent PDR chemistry \citep{Bergin1997,Harada2015}.  An advantage of the time-dependent PDR picture over the CRDR picture is that the \NNHp\ and \HCOp\ abundances increase in later phases than \Cn\ and CN, being consistent with the PCA results.

The kinetic energy and mass of the \CI-bright ring are marginally consistent with those of the material that can be collected by Sgr~A~East.
The mass $M$ is estimated to be $3\times10^4\ \Msol$ from the excess \CI\ emission $\dCI_{10}$ integrated over the \CI-bright ring, assuming that the \CI-ring consists of gas in the \Cn-dominant phase.  
The kinetic energy $K$ is given as $K = \frac{1}{2}M \sigma_v^2 = 0.7\times10^{50}\ \erg$, where $\sigma_v \sim 15\ \kmps$ is 3-D velocity dispersion of the entire ring.
The estimated $K$ is an order lower than the typical energy of an SNR explosion and consistent with the kinetic energy injected into the W49 GMC via interaction with SNR \citep{Sashida2013}.
The mass of the low-density material that existed inside the \CI-bright ring before the SN explosion is $\phi\frac{4\pi}{3}R_{\rm ring}^3\rho_{\rm d}$, where $R=6.5\ \pc$ is the major radius of the ring, $\rho_{\rm d}$ is the mass volume density, and $\phi\sim\frac{2}{3}$ is the initial volume filling factor of the diffuse molecular gas \citep{Takeshi2019,Oka2020}.
By using the present value for $\nHH$ in the diffuse molecular cloud in Sgr~A, 150--350~$\pcc$ \citep{Goto2008}, we obtain $M = \left(0.8\text{--}2\right)\times 10^4\ \Msol$.  
As it is unlikely that the diffuse ionized and atomic gas have a higher density than the diffuse clouds, the mass of those low-density materials inside the \CI-bright ring is marginally comparable to the mass of the \CI-bright ring.

In the previous paper \citep{Tanaka2011}, we hypothesized that chemically young low-density gas has been supplied to the Sgr~A~complex by the gas flow facilitated by the nested bar potential \citep{Namekata2009}.   
However, we were not able to identify such streaming motion in the PV structure of the \CI-excess map.  

\subsubsection{Origin of the \CI-bright ring}
To summarize the discussion above, the CR dissociation is one of the plausible origins of the enhanced \Cn\ abundance in the \CI-bright ring, as it reasonably explains the observed ring morphology encircling the SNR shell and the enrichment of CN.
The time-dependent PDR chemistry in the primitive molecular gas collected by the SNR is a possible alternative hypothesis.
Other scenarios hypothesized in the previous paper \citep{Tanaka2011} have difficulties in explaining the CN enrichment or the morphology of the \CI-bright region.

We do not conclude which of the CRDR and time-dependent chemistry hypotheses is more likely in this paper.
The CR dissociation picture has the advantage that it explains the origin of all \CI-enhanced regions by assuming a single CR source, namely, Sgr~A~East, because the projected positions of the CND and CO0.02 are also within the estimated range of the CR diffusion distance.  
On the other hand, the time-dependent PDR picture better explains the absence of the enhancement of the \NNHp\ and \HCOpt\ intensities in the \CI-enhanced region.
To conclude about the origin of the \CI-enhanced region, additional analysis such as more direct measurements of \zCR\ values in dense gas (e.g.,  \citealt{Ceccarelli2011,Vaupre2014}) would be crucial.

\subsection{\Cn-enrichment in the CMZ\label{discussion:overall}}

\begin{figure}
%\ifdraft
\epsscale{1.2}
%\else\epsscale{1.2}\fi
    \plotone{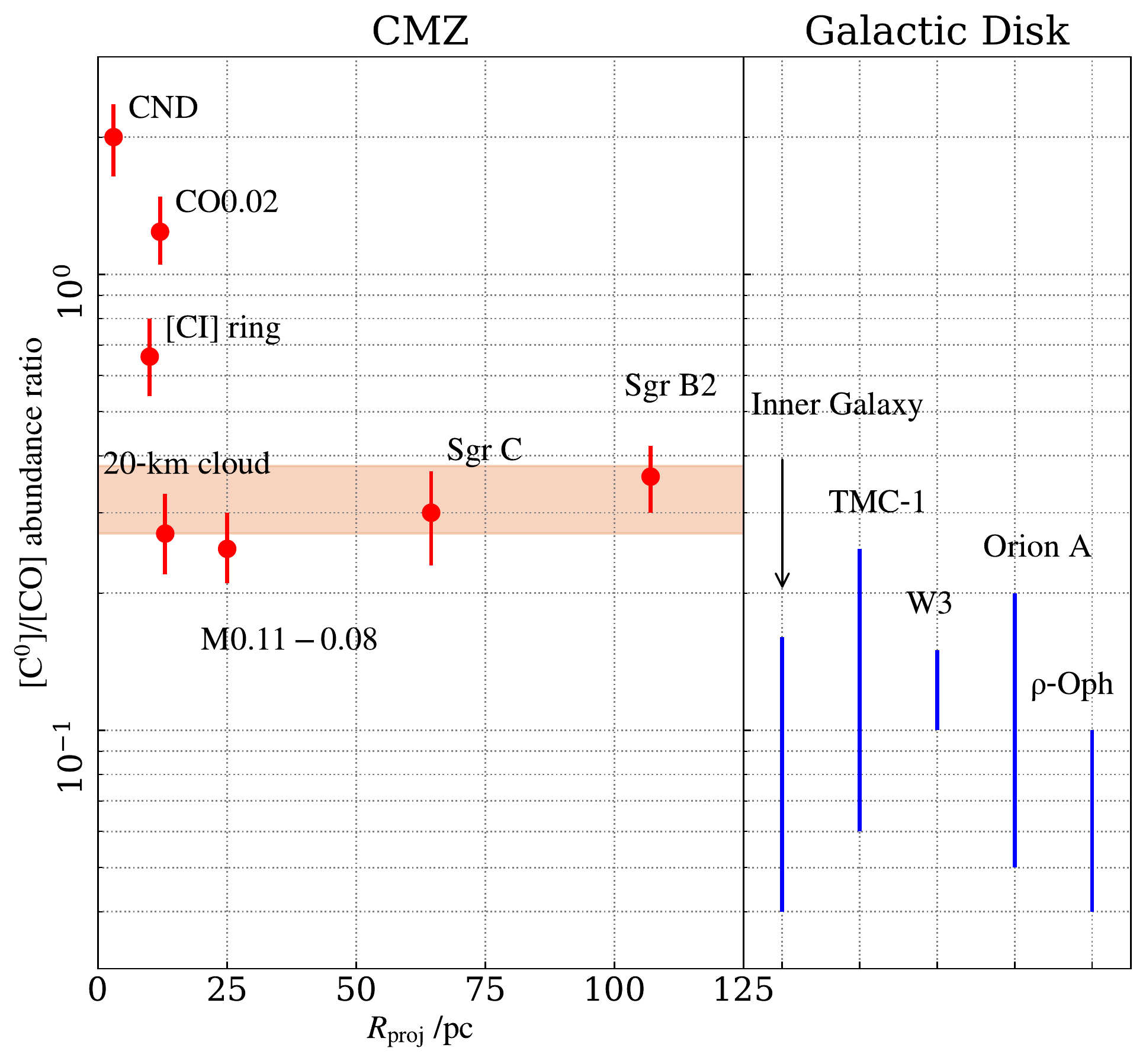}
    \caption{{\NN(\Cn)/\NN(\CO)} abundance ratios in molecular clouds in the CMZ and the Galactic disk region.  The CMZ clouds are plotted against the projected distance from \sgras. {The CMZ averaged value is indicated by the red hatched region.}
    {The \xCn\ abundance ratios in the Galactic disk clouds, TMC-1, W3, Orion~A, and $\rho$-Oph are taken from \cite{Maezawa1999}, \cite{Sakai2006}, \cite{Ikeda2002}, and \cite{Kamegai2003}, respectively.  The ratio in the inner Galaxy is calculated from the \CI~\CIa\ and CO~\JJ{1}{0}\ intensities taken from \cite{Fixsen1999}.  }
    \label{fig:r_CI}}
\end{figure}

\newcommand\Rproj{\ifmmode R_{\rm proj}\else$R_{\rm proj}$\fi}
\newcommand\Xco{\ifmmode X_{\rm CO}\else$X_{\rm CO}$\fi}
%\begin{revision}
Figure \ref{fig:r_CI} compares the {\NN(\Cn)/\NN(\CO)} abundance ratios of the CMZ clouds with those in the Galactic disk region taken from the literature \citep{Fixsen1999,Maezawa1999,Sakai2006,Ikeda2002,Kamegai2003}.
The abundance ratios in the solar neighborhood clouds (TMC-1, W3, Orion-A, and  $\rho$-Oph) are all measured using the \CI~\CIa\ and \JJ{1}{0}\ line of CO and \COt, and hence represent the ratios in low-density gas components similar to the low-excitation component of the CMZ.
The Inner Galaxy value is calculated from the \CI~\CIa\ and $^{12}\CO$~\JJ{1}{0}\ intensities taken from the {\it COBE}/FIRAS all-sky survey data \citep{Fixsen1999},
by assuming a \CI~\CIa\ excitation temperature of 20 K and a CO~\JJ{1}{0}\ intensity to \NHH\ conversion factor of $2\times10^{20}\ \psc/({\kelvin\cdot\kmps})$.
The {\NN(\Cn)/\NN(\CO)} abundance ratios in the solar-neighborhood clouds and the inner Galaxy are within the range of a few 0.01--0.2. 
%\end{revision}
The ratios in the {CMZ clouds without \CI-enhancement} are a factor of 2--3 above the Galactic disk clouds.
The \xCn\ abundance ratio increases to 0.6--1 in the \Cn-rich clouds inside $\sim 10$ pc projected galactocentric radius $\Rproj$.  
The highest \Cn\ abundance is measured in the CND, where {\NN(\Cn)/\NN(\CO) ratio} exceeds unity; however, 
a smooth spatial gradient of the \NN(\Cn) abundance does not present in $\Rproj \gtrsim 10~\pc$, where the {\NN(\Cn)/\NN(\CO)} abundance ratio is uniformly $\sim 0.3$.
No systematic difference is found between the clouds with \ion{H}{2} regions and those without in both the CMZ and the solar-neighborhood.

We may consider the same origin for the overall \Cn\ enrichment in the CMZ as that for 
the \CI-bright ring, namely, CR dissociation or time-dependent chemistry.
The \zCR\ value in the interior of the Sgr~B2 complex is measured to be $\sim 10^{-16}\ \ps$ through observations of emission lines of $\rm H_3O^+$ and complex organic molecules \citep{Tak2006,Willis2020}, despite the absence of nearby remarkable CR sources such as SNRs;
the high \Cn\ abundance may indicate that \zCR\ is generally enhanced from the Galactic disk value in dense gas, in accordance with the high \zCR\ in diffuse clouds measured with \HHHp\ absorption study \citep{OkaTakeshi2005,Goto2008,Takeshi2019}. 
In terms of the time-dependent chemistry, the high \xCn\ abundance ratio is interpreted as faster lifecycle of the CMZ clouds than in the Galactic disk.
Indeed, theoretical calculation shows short cloud lifetimes of 1.6--3.9 Myr in the CMZ \citep{Jeffreson2018}, which are a factor of $>2$ shorter than those in the larger radii and comparable to the chemical time scale of the $\Cn$ to CO conversion.
\cite{Harada2019} argue that the effective chemical age is further limited by the turbulent crossing time due to the continuous turbulent mixing of the cloud interior and PDR at the cloud surface. 
The size-line~width relationship of the CMZ cloud has a factor of 5 higher velocity scale than that in the Galactic disk \citep{Tsuboi1999,Shetty2012,Tanaka2020}, yielding generally shorter turbulent crossing time and hence a higher \Cn\ abundance.
In either case of the high \zCR\ and the fast cloud lifecycle, 
the \Cn\ abundance decreases with increasing \nHH, being consistent with the low \Cn\ abundance in the high excitation component.

\section{SUMMARY\label{section:summary}}

We presented the \CI~\CIa\ mapping of the CMZ extended from the previous study  \citep{Tanaka2011}\ and searched for \CI-bright regions with enhanced \CI\ brightness with improved sensitivity and wider spatial coverage.
The spatial variation in the \Xcn\ abundance ratio is investigated by performing non-LTE analysis using multiline \COt\ and CO lines along with the \CI\ map.
Based on the \Xcn\ abundance measurement and PCA involving 17 molecular/atomic maps in the Sgr~A complex, we discussed the origin of the \Cn-rich regions, in particular, that of the \CI-bright ring surrounding the Sgr~A~East SNR.
The main results are summarized below:

\begin{itemize}
    \item The tight intensity correlation of the \CI\ with the \COt~\JJ{1}{0} and \JJ{2}{1}\ lines in the PPV space and the PCA results indicates that \CI~\CIa\ predominantly traces the low-excitation component of the CMZ molecular gas with \Tkin = 20--50~\kelvin\ and \nHH\ of $\sim10^3$~\pcc.  The contribution from the high-excitation component with \Tkin\ of $\sim 100$~K and \nHH\ of $\sim 10^{4\text{--}5}\ \pcc$ to the \CI\ emission is less than that to the low-$J$ \COt\ lines.  
    
    \item The bulk component of the CMZ gas has overall uniform intensity ratios of the \CI\ to the \COt~\JJ{1}{0}\ and \JJ{2}{1}\ lines, which are 0.43 and 0.51, respectively.  These ratios translate into the \xCn\ abundance ratio of 0.3.   The Sgr~B2 and C complexes, 20-\kmps\ cloud, and M0.11$-0.08$ belong to this component with the normal \Cn\ abundance.
    
    \item The regions with substantially enhanced \CI-brightness are present in the CND, CO0.02$-0.02$, and the high-velocity portion of the 50-\kmps\ cloud.  They were all reported in \cite{Tanaka2011};  no new \CI-bright regions were discovered in the extended mapping regions.   The \CI-bright clouds have the \CI/\COt~\JJ{1}{0}\ and \CI/\COt~\JJ{2}{1}\ ratios that are more than twice that of the {clouds without \CI-enhancement}.   The non-LTE analysis has confirmed that this enhanced \CI/\COt\ intensity ratio is owing to enhanced \Cn\ abundance, not a result of higher \COt\ excitation temperatures.   The highest \Cn\ abundance is measured for the CND, where $\xCn\sim 2$.
    
    \item The \CI-bright cloud in the high-velocity portion of the 50-\kmps\ cloud has a distinct ring-like morphology encircling the radio shell of the SNR Sgr~A~East.  The PV structure of this \CI-bright ring is apparently decoupled from the global velocity gradient of the Sgr~A complex, indicative of interaction with the SNR. The \Xcn\ abundance ratio of the \CI-bright ring is $\ge 0.7$.
%    which is possibly contaminated by the overlapping clouds with normal \Cn\ abundance.
    
    \item In the PCA results, the relative enhancement of the \CI\ intensity to the low-$J$ \COt\ intensities are represented by the PC4 and 5.  The only other line that has large positive PC4 and 5 loading is CN~\JN{1}{3/2}{0}{1/2}.   Neither shock tracers (SiO, HCN, CS, and \HCCCN) nor quiescent-gas tracers (HNC, \HCOp, and \NNHp) do not increase in the \CI-bright regions.  
    
    \item The morphological correlation of the \CI-bright ring and the outer edge of the SNR shell indicates that the increase in the \Cn\ and CN abundances is owing to enhanced \zCR\ in the SNR-shocked clouds.  The same degree of the \zCR\ enhancement ($\zCR\sim 10^{15}\ \mathrm{s}^{-1}$) measured in the dense molecular cloud interacting with the IC443 SNR explains the observed high \Cn\ and CN abundances.   The radial extension of the \CI-bright regions ($< 5$ pc away from the Sgr~A~east shell) is consistent with the age of the SNR ($(1\text{--}10)\times10^4\ \yr$) and the diffusion coefficient of the low-energy CR proton in the $B \sim 100\ \mu\mathrm{G}$ magnetic field typical in the CMZ.
    
    \item The high \Cn\ and CN abundances in the ring are alternatively understood as the \Cn-rich phase in the time-dependent chemistry, which could appear in primitive molecular gas within $\sim 0.8\ \Myr$ after formation via compression by the expanding SNR.  The kinetic energy calculated assuming an expanding motion ($0.7\times10^{50}\ \ergs$) and the mass ($3\times10^4\ \Msol$) are marginally consistent with the gas collected by the SNR.
    
    \item The \xCn\ abundance ratio averaged over the low-excitation component in the CMZ, 0.3, is 2--3 times those measured for solar neighborhood clouds.   The overall enhanced \zCR\ and shorter cloud lifetime than in the Galactic disk region may explain the enhanced \Cn\ abundance in the CMZ.   The \Cn\ abundance in the high-excitation component is not directly measured; however, the PCA results suggest a dearth of \Cn\ in the high-excitation component, whose \xCn\ abundance ratio is likely $<0.1$.
    
    \item The steady-state PDR, XDR, and shock-induced chemistry are not favored as the origin of the \CI-bright ring by the observed morphology and chemical characteristics of the ring.  The inflowing low-density material from outside the Sgr~A complex hypothesized as the source of the \Cn-rich gas in \cite{Tanaka2011} is not identified in the new \CI\ map.
    
\end{itemize}

%\bibliographystyle{apj}
%\bibliography{mendeley,local}
 \newcommand{\noop}[1]{}

\appendix

\section{The \COt, CN, and \methanol\ Data \label{appendix:OtherMaps}}

\begin{figure}
  \begin{center}
  \epsscale{0.65}
  \plotone{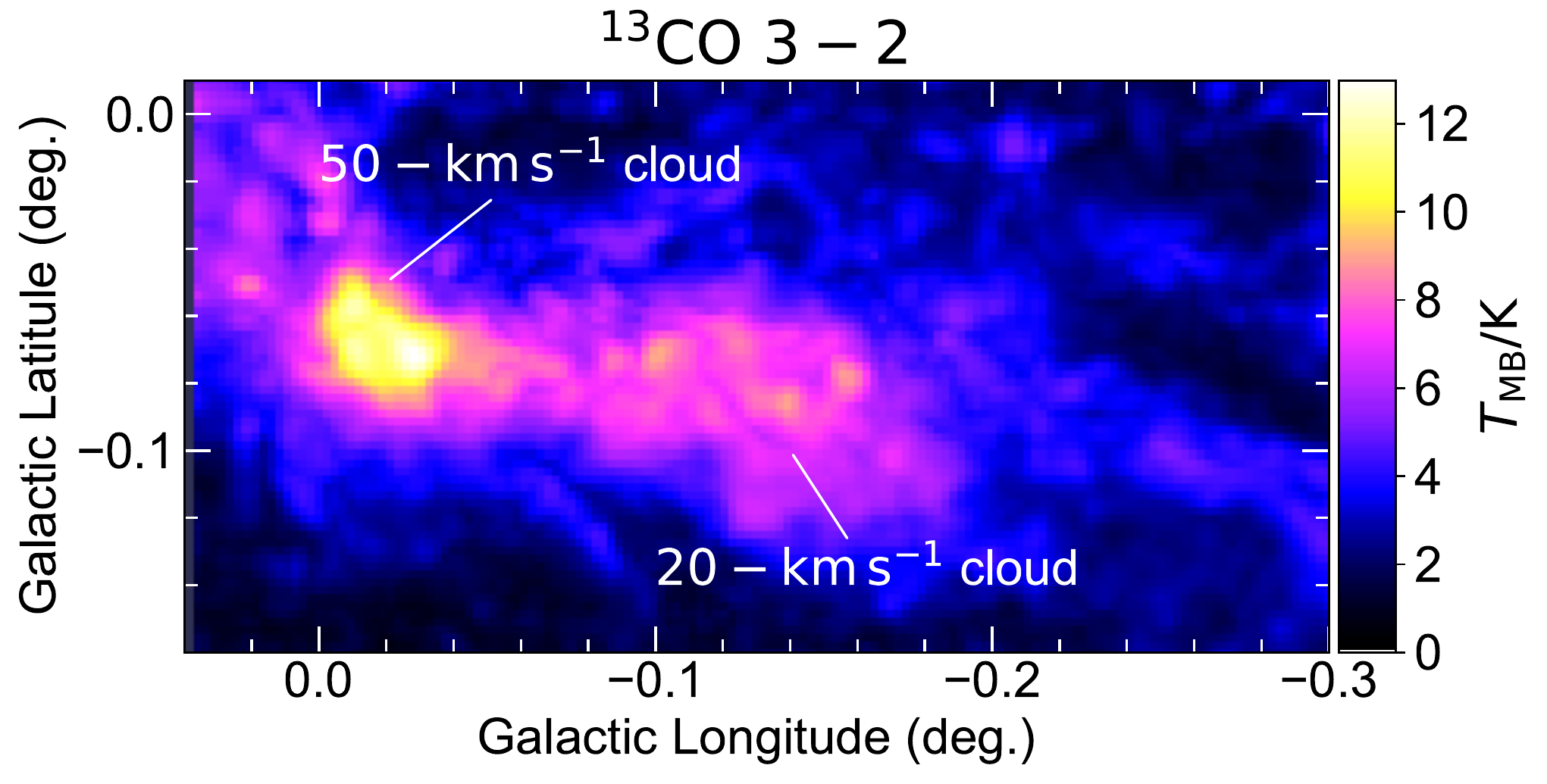}
  \end{center}
  \caption{\COt~\JJ{3}{2}\ peak-intensity map obtained with the ASTE 350-GHz band observations.  The peak intensities are calculated in 10-\kmps\ velocity bins.
    \label{fig:map_13CO32}}
%\end{figure}

%\begin{figure}[ttt]
  \begin{center}
    \epsscale{1.15}
    \plotone{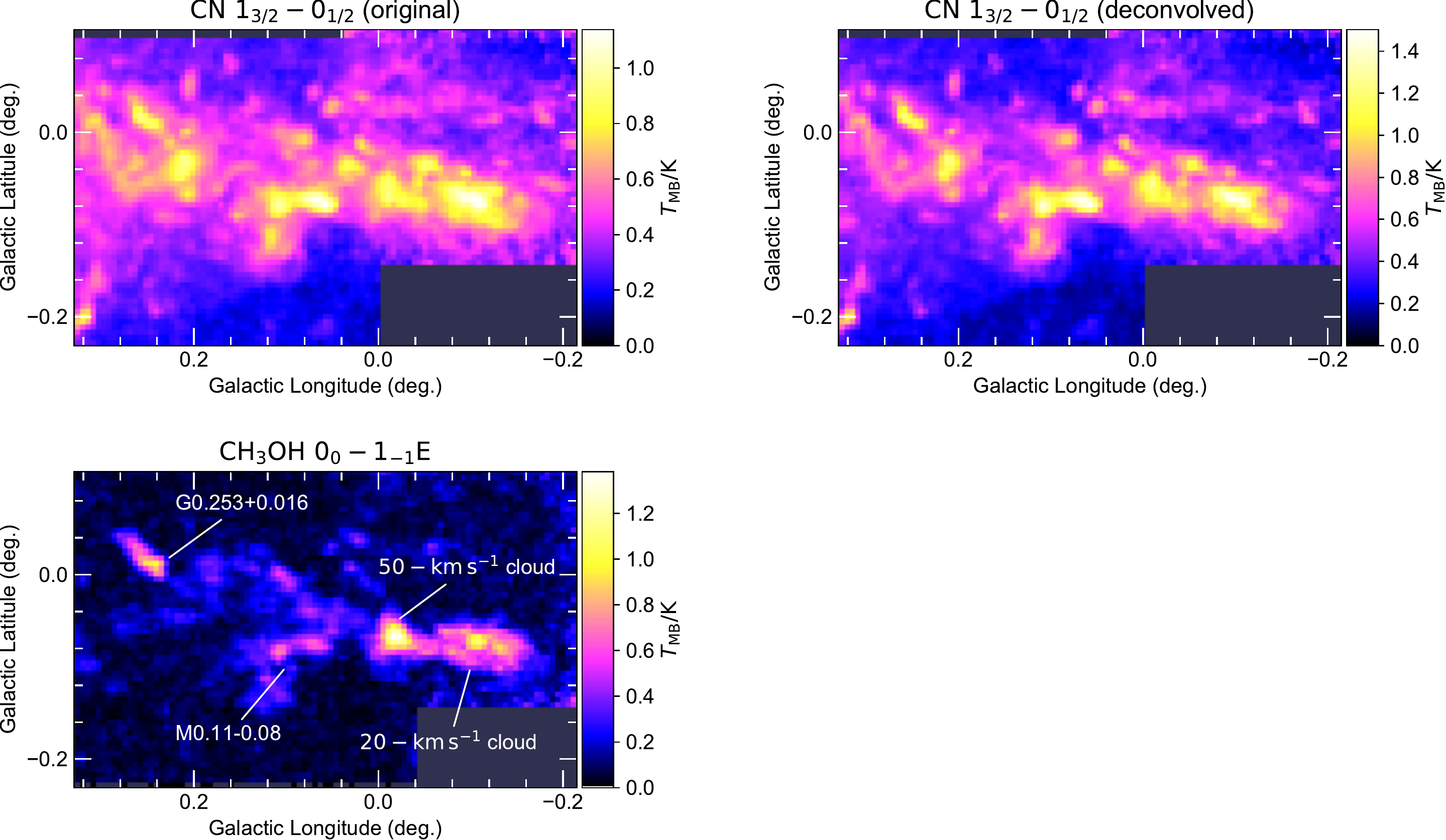}
    \caption{CN \NJ{1}{3/2}{0}{1/2}\ and
      \methanol\ $0_0$--$1_{-1}$E peak-intensity maps (top and bottom rows, respectively) obtained with the NRO 45-m observations. The peak intensities are calculated in the 10-\kmps\ velocity bins. Both the raw and the hyperfine-deconvolved maps are shown for CN.
    \label{fig:map_NRO}}
  \end{center}
\end{figure}

Figure \ref{fig:map_13CO32} shows the peak-intensity map of the \COt~\JJ{3}{2}\ line obtained with the ASTE-10m telescope.

Figure \ref{fig:map_NRO} shows the peak-intensity maps of the CN~\JN{1}{3/2}{0}{1/2}\ and \methanol~$0_0\text{--}1_{-1}$E lines taken with the NRO 45-m telescope.  For the CN map, we show both the original and the hyperfine-deconvolved data.  The deconvolved peak-intensity has a distribution with higher contrast.   The details of the hyperfine deconvolution are given in Appendix \ref{appendix:CNdeconv}.

\section{Hyperfine Deconvolution using the Fourier Quotient Method \label{appendix:CNdeconv}}
\newcommand\profhf{\ifmmode\phi_\mathrm{HF}(v)\else$\phi_\mathrm{HF}(v)$\fi}
\newcommand\proforg{\ifmmode\phi(v)\else$\phi(v)$\fi}
\newcommand\fker{\ifmmode K(v)\else$K(v)$\fi}

\newcommand\kphi{\tilde{\phi}}
\newcommand\fprofhf{\ifmmode\kphi_\mathrm{HF}(k)\else$\kphi_\mathrm{HF}(k)$\fi}
\newcommand\fproforg{\ifmmode\kphi(k)\else$\kphi(k)$\fi}
\newcommand\ffker{\ifmmode \tilde{K}(k)\else$\tilde{K}(k)$\fi}

In the optically thin limit, a hyperfine-splitting profile \profhf\ is expressed as a convolution of the original profile \proforg\ and a kernel function \fker ;
\begin{eqnarray}
\profhf = (\phi * K)(v)\label{eqn:hfconv}
\end{eqnarray}
In the Fourier domain, convolution is converted into the product of the Fourier transforms (FTs) of the functions.
Hence, the hyperfine deconvolution is performed by calculating the quotient between the FTs of \profhf\ and \fker ;
\begin{eqnarray}
\fproforg = \frac{\fprofhf}{\ffker}, \label{eqn:fq}
\end{eqnarray}
 with \fproforg, \fprofhf, and \ffker\ being the FTs of \proforg, \profhf, and \fker, respectively.  
 The deconvolved profile is obtained as an inverse-FT of \fproforg.
 As obvious from Equation \ref{eqn:fq}, this method is applicable on the condition that $|\ffker| > 0$ for all $k$.
The kernel function \fker\ is given as 
\begin{eqnarray}
\fker = \frac{\sum f_i\cdot\delta(v-v_i)}{\sum f_i},
\end{eqnarray}
where $f_i$ and $v_i$ are the relative intensity and frequency offset expressed in the Doppler velocity of the $i^\mathrm{th}$ hyperfine components;  their values for CN \JN{1}{3/2}{0}{1/2}\ are shown in Table \ref{table:CNdeconv}.   
The condition $|\ffker| > 0$ is fulfilled for the case of the CN line.
In the actual calculation, the Fourier quotient (Equation \ref{eqn:fq}) is applied to the channels with sufficiently high S/N ratios in the Fourier domain to avoid artifacts from noise.
The cutoff level is chosen as 2-$\sigma$ for our CN data.  
The deconvolution is performed position-by-position using the noise levels of the individual spectra. 

\begin{deluxetable}{lcc}
\tabletypesize{\small}
\tablewidth{18cm}
\tablecolumns{3}
\tablecaption{Hyperfine Components of the CN $N_J$ = $1_{3/2}$--$0_{1/2}$ line \label{table:CNdeconv}}
\tablehead{
\colhead{} & \colhead{$f_i$} & \colhead{$v_i$} \\ 
\colhead{$\hspace{70pt}$} & \colhead{$\hspace{70pt}$} & \colhead{$\hspace{30pt}(\kmps)\hspace{30pt}$}
}
\startdata
$F$=${1/2}$--${3/2}$ & 0.037 & $-77.8$\\
$F$=${3/2}$--${3/2}$ & 0.296 & $-47.4$\\
$F$=${1/2}$--${1/2}$ & 0.296 & $-22.9$\\
$F$=${5/2}$--${3/2}$ & 1.000 & \phs\phn$0.0$\\
$F$=${3/2}$--${1/2}$ & 0.370 & \phs\phn$7.5$ 
\enddata

\end{deluxetable}

\begin{figure}
    \epsscale{1.}
    \plotone{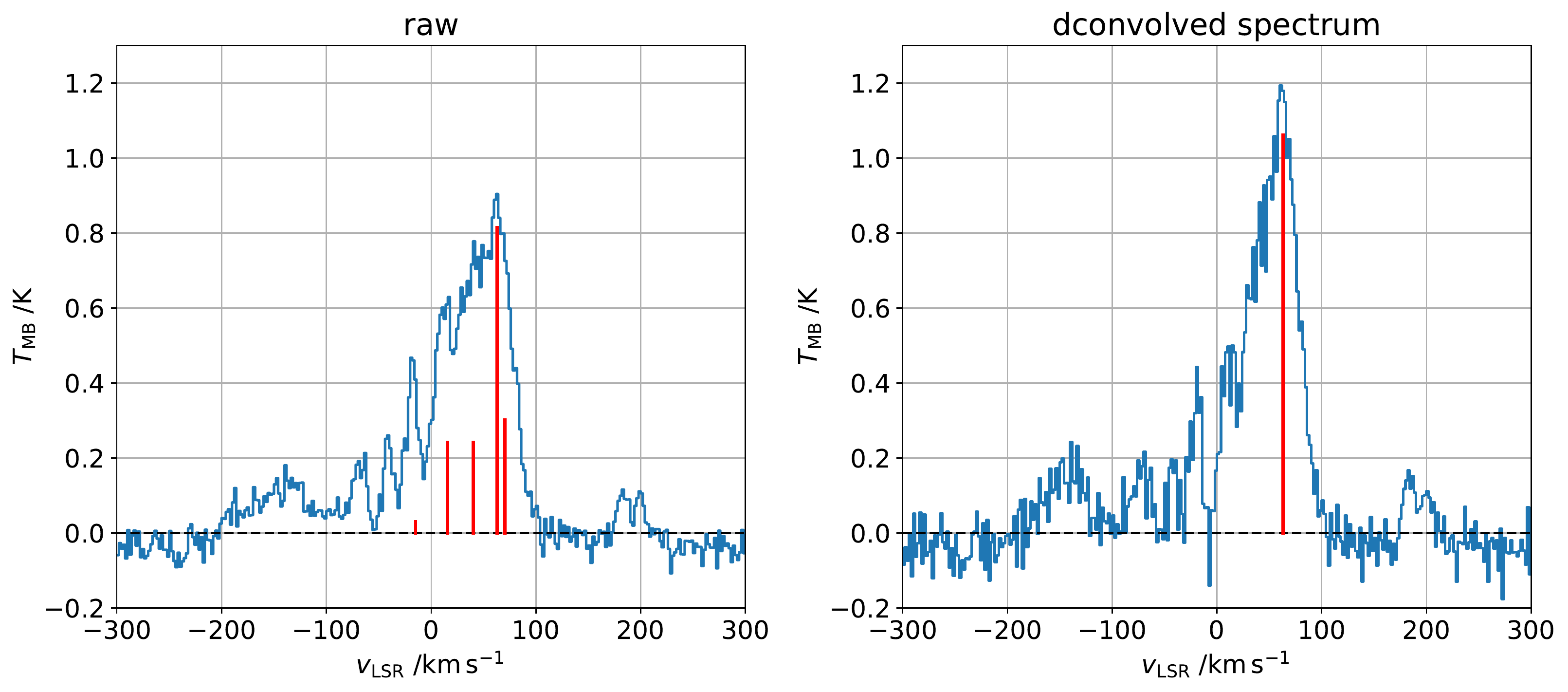}
    \caption{CN \JN{1}{3/2}{0}{1/2}\ spectra averaged over a $60''$ diameter circle around the peak position of the 50-\kmps\ cloud ($\gl, \gb = -0.011^\circ, -0.062^\circ$) before and after hyperfine deconvolution (left and right panels, respectively).   The relative intensities and velocity offsets of the hyperfine components are schematically denoted by vertical bars in the left panel.
    \label{fig:FQspec} }
\end{figure}

Figure \ref{fig:FQspec}\ shows the original and deconvolved CN spectra averaged over a $60''$ diameter circle around the peak position of the 50-\kmps\ cloud.
As the velocity intervals among the hyperfine components are less than the velocity width of the original spectrum, the hyperfine splitting is observed as line-broadening in the original profile.  
A sharper profile is successfully obtained after the Fourier quotient is applied. 
In this example, the hyperfine deconvolution has increased the peak intensity by a factor of 1.3, while the noise level has also increased by a factor of 1.5 as a side effect;  hence, the dynamic range of the spectrum data is approximately conserved before and after the deconvolution.

%\begin{revision}
\section{Principal Component Analysis\label{appendix:PCA}}
\begin{figure}
    \epsscale{.65}
    \plotone{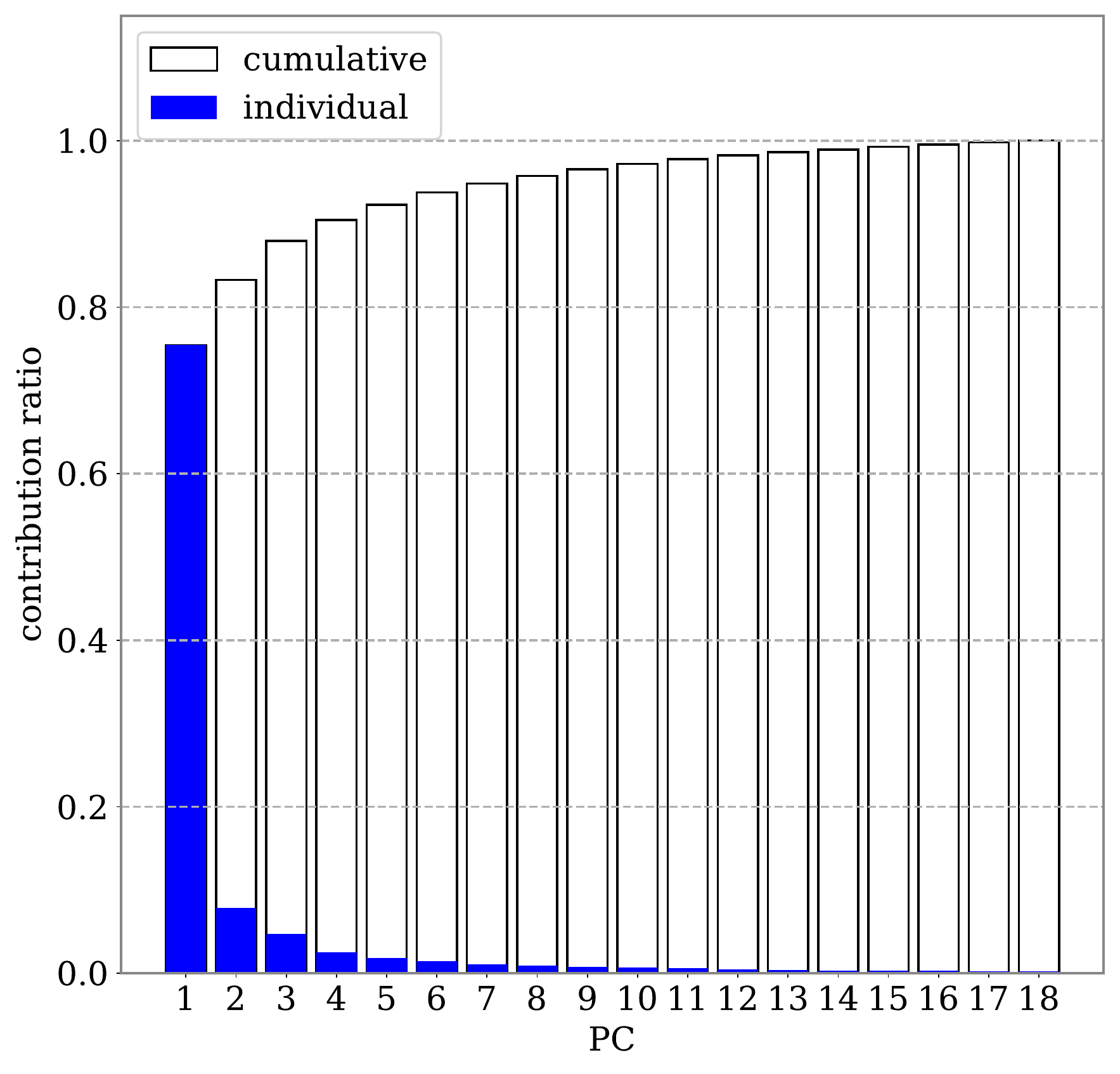}
    \caption{Contribution ratios and cumulative contribution ratios in the PCA in \S\ref{section:analysis:PCA}.\label{fig:pccontrib}}
\end{figure}

%Let $\vpi$ be a $p$-dimensional vector whose elements are the input line intensities at the $i^\mathrm{th}$ voxel ($i = 1, 2, ..., \Np$) and
Principal component analysis is a method for evaluating similarity/dissimilarity between different images based on the diagonalization of the covariance matrix.
Let $P = (p_{i,j})$ a $M\times N$ matrix ($M>N$) whose elements are the $j^\mathrm{th}$ line intensity at the $i^\mathrm{th}$ voxel.
%$P$ a matrix whose columns are $\vp_1, \vp_2, ..., \vp_\Np$. 
For simplicity, we assume that all line intensities are normalized so that their averages and variances are 0 and 1, respectively. 
In general, different line intensities are intercorrelated, i.e, the covariance matrix $S \equiv \frac{1}{M}\tr{P}P$ has non-zero off-diagonal elements.
PCA finds the linear combinations of the original parameters that are orthogonal and hence better describe the variation of the parameters.
Such linear combinations are obtained by diagonalizing covariance matrix $S$; 
\begin{eqnarray}
\tr{V}SV = \frac{1}{M}\tr{(PV)}PV = \left(
    \begin{array}{cccc}
      \lambda_1 &           &        &        \\
                & \lambda_2 &        &        \\
                &           & \ddots &        \\
                &           &        & \lambda_N 
    \end{array}\right)
 , \label{eqn:axisconv}
\end{eqnarray}
where $V = (v_{i,j})$ is a matrix whose columns are the eigenvectors $\vv_i$ ($i$=1,2,...,$N$) and $\lambda_1, \lambda_2, ..., \lambda_N$ are the eigenvalues.
$\vv_i$ and $\lambda_i$ are indexed in the decreasing order of the eigenvalues.
%Matrix $V$ defines the axis conversion from the original parameter space (i.e., the normalized line intensities) to a new orthogonal axes called principal components (PCs).  
%The new coordinate system defined by $V$ is the called principal components (PCs).
The new axis defined by eigenvector $\vv_i$ is called the  $i^\mathrm{th}$ principal component (PC), denoted as PC$i$.
PC1 is the direction along which the variance is the largest, and PC2 the second largest, and so on.
%The direction of the largest dispersion is defined as PC1, the second largest as PC2, and so on.
%The dispersion along PC$i$ is equal to $\lambda_i$.
%Equation \ref{eqn:axisconv} means that the variance along PC$i$ is equal to $\lambda_i$ and covariance between difference PCs are 0; in this sense, PCs can be regarded as the fundamental variables in the data.
By defining the PC score vector $\vqi$ to be the $i^\mathrm{th}$ column of matrix PV, Equation \ref{eqn:axisconv} is rewritten as $\tr{\vq_i}\cdot\vq_j = M \lambda_i\delta_{i,j}$, i.e., the PCs are orthogonal to each other and the variance along the PC$i$ axis is equal to the eigenvalue $\lambda_i$.
The relative importance of PC$i$ in the data space is measured by the contribution ratio defined as $\mathrm{CR}_{i}\equiv\lambda_i/\sum_{i=1}^{N}\lambda_i = \lambda_i/N$.
The contribution ratios and cumulative contribution ratios ($\sum_{k\leq i} \mathrm{CR}_{k}$) in the analysis described in \S\ref{section:analysis:PCA} are shown in Figure \ref{fig:pccontrib}.

The parameter vector $\pvi$ defined as the $i^\mathrm{th}$ column of matrix $P$, i.e., a vector of the voxel values of the $i^\mathrm{th}$ line intensity, is related to the PC score vector \qvi\ by 
\begin{eqnarray}
\pvi & = & \left(\qv_1, \qv_2, ..., \qv_N \right)\left(
\begin{array}{c}
v_{1,i} \\
v_{2,i} \\
\vdots \\
v_{N,i} 
\end{array}
\right).
\end{eqnarray}
Thus the information of the variation of the $i^\mathrm{th}$ line intensity over the $M$ voxels is reduced into $N$ parameters $v_{1,i}, v_{2,i}, ..., v_{N,i}$. 
The PC loading diagrams presented in Figure \ref{fig:PCloadingVectors} show the elements of the PC eigenvectors $v_{i,j}$ for the first 5 PCs.
We can evaluate the similarity or dissimilarity between the line intensity maps according to their loci on the loading diagram.
%We can identify lines with similar distributions according to their loci in the loading diagram.

Figure \ref{fig:pc45score_synth} shows the \CI\ maps reconstructed using selected PCs (PC${i_1}$, PC${i_2}$, ...): 
\begin{eqnarray}
\tilde{\vp} (i_1, i_2, ...) & = & \sum_{i\in\{i_1, i_2, ...\}} v_{i, k}\cdot \qv_i \label{eqn:pccompmap} , 
\end{eqnarray}
where index $k$ corresponds to \CI.
The PC4+PC5 map (i.e., $\tilde{\pv} (4,5)$) in Figure \ref{fig:pc45score_synth} shows a ring structure similar to that in the \dCI\ maps (Figure \ref{fig:dCI}), indicating that the \CI-enhancement is represented by PC4 and PC5.

%, which plots the elements of the PC eigenvectors $v_{i,j}$.

%By defining the 

%Equation \ref{eqn:axisconv} is written as $\tr{\vqi}\cdot{\vqi} = \lambda_i\delta_{i,j}$.
%In the PC space, the line intensities at the $i^\mathrm{th}$ voxel are expressed in the PC scores defined as $\vqi = \tr{V}\vpi$.
% in the PC space.
%The importance of the individual PCs is measured by the contribution ratio defined by $\lambda_i/\sum_{j=1}^{N}\lambda_j$.
%Hence, the information on the variation of the $i^\mathrm{th}$ line intensity over the $\Np$ voxels is %reduced into a single $N$-dimensional vector defined by the $i^\mathrm{th}$ row of $V$.

%The line intensity distributions are analyzed 
%Hence, we can compare the line intensity distributions on the PC loading plots, on which 

%The dispersion along PC$i$ is equal to the eigenvalue $\lambda_i$.  
%Figure \ref{fig:axisConversion} schematically illustrates the conversion from the original parameter space to the PC space for the case of $N=2 $.

%In the PC space, the data points are expressed in the PC scores defined as $\vqi = \tr{V}\vpi$. % instead of the raw line intensities.

%; i.e., different PCs are orthogonal to each other and the dispersion along the PC$i$ axis is equal to $\lambda_i$.   In this sense, the PC scores can be regarded as the fundamental values in the input data set.  

%\end{revision}

\end{document}